\documentclass{emulateapj}

\begin{document}

\title{Simulations on a Moving Mesh: \\The Clustered Formation of Population~III Protostars}

\author{Thomas~H.~Greif\altaffilmark{1}, Volker~Springel\altaffilmark{2,3}, Simon~D.~M.~White\altaffilmark{1}, Simon~C.~O.~Glover\altaffilmark{4}, Paul~C.~Clark\altaffilmark{4}, Rowan~J.~Smith\altaffilmark{4}, Ralf~S.~Klessen\altaffilmark{4} \& Volker~Bromm\altaffilmark{5}}

\altaffiltext{1}{Max-Planck-Institut f\"{u}r Astrophysik, Karl-Schwarzschild-Stra\ss e 1, 85740 Garching bei M\"{u}nchen, Germany}
\altaffiltext{2}{Heidelberg Institute for Theoretical Studies, Schloss-Wolfsbrunnenweg 35, 69118 Heidelberg, Germany}
\altaffiltext{3}{Zentrum f\"{u}r Astronomie der Universit\"{a}t Heidelberg, Astronomisches Recheninstitut, M\"{o}nchhofstr. 12-14, 69120 Heidelberg, Germany}
\altaffiltext{4}{Zentrum f\"{u}r Astronomie der Universit\"{a}t Heidelberg, Institut f\"{u}r Theoretische Astrophysik, Albert-Ueberle-Stra\ss e 2, 69120 Heidelberg, Germany}
\altaffiltext{5}{Department of Astronomy and Texas Cosmology Center, University of Texas, Austin, TX 78712, USA}

\begin{abstract}
The cosmic dark ages ended a few hundred million years after the Big Bang, when the first stars began to fill the universe with new light. It has generally been argued that these stars formed in isolation and were extremely massive -- perhaps $100$ times as massive as the Sun. In a recent study, Clark and collaborators showed that this picture requires revision. They demonstrated that the accretion disks that build up around Population~III stars are strongly susceptible to fragmentation and that the first stars should therefore form in clusters rather than in isolation. We here use a series of high-resolution hydrodynamical simulations performed with the moving mesh code {\small AREPO} to follow up on this proposal and to study the influence of environmental parameters on the level of fragmentation. We model the collapse of five independent minihalos from cosmological initial conditions, through the runaway condensation of their central gas clouds, to the formation of the first protostar, and beyond for a further $1000$~years. During this latter accretion phase, we represent the optically thick regions of protostars by sink particles. Gas accumulates rapidly in the circumstellar disk around the first protostar, fragmenting vigorously to produce a small group of protostars. After an initial burst, gravitational instability recurs periodically, forming additional protostars with masses ranging from $\sim 0.1$ to $10\,{\rm M}_\odot$. Although the shape, multiplicity, and normalization of the protostellar mass function depend on the details of the sink-particle algorithm, fragmentation into protostars with diverse masses occurs in all cases, confirming earlier reports of Population~III stars forming in clusters. Depending on the efficiency of later accretion and merging, Population~III stars may enter the main sequence in clusters and with much more diverse masses than are commonly assumed.
\end{abstract}

\keywords{cosmology: theory --- early universe --- hydrodynamics --- methods: numerical --- stars: formation}

\maketitle

\section{Introduction}

Over the past decade, a consensus has emerged on how the first, so-called Population~III (Pop~III), stars may have formed out of the hydrogen and helium forged in the Big Bang \citep{bl01,bl04a,bromm09}. Non-linear structure formation from an initially nearly featureless universe began when dark matter minihalos with masses of order $10^6\,M_\odot$ collapsed at redshifts $z\sim 20-50$, confining gas within their gravitational potential wells \citep{htl96,tegmark97}. Cooling through ro-vibrational transitions of newly formed molecular hydrogen then triggered runaway collapse at the center of the halo, resulting in the formation of a protostar with density more than twenty orders of magnitude higher than the cosmic mean \citep{yoh08}. The envelope of the accreting protostar remained hot due to the lack of cooling by metals and dust, so that accretion rates were on average higher than in star-formation regions today \citep{mk04,mo07,zy07}. With a few exceptions \citep{tao09}, simulations of this initial collapse phase have shown no fragmentation \citep{abn02,bcl02,bl04b,yoshida06b,on07,yoh08}, leading to the conclusion that the first stars formed in isolation and were extremely massive.

\begin{table*}
\begin{center}
\caption{Simulation parameters}
\begin{tabular}{ccccccccccc} \hline \hline
Simulation & Size [kpc] & Particles & $M_{\rm dm}~[{\rm M}_\odot]$ & $M_{\rm dm, ref}~[{\rm M}_\odot]$ & $M_{\rm gas}~[{\rm M}_\odot]$ & $\sigma_{8}$ & $M_{\rm vir}~[{\rm M}_\odot]$ & $r_{\rm vir}~[{\rm pc}]$ & $\lambda$ & $z_{\rm coll}$ \\ \hline
MH-1 & $1000$ & $512^3$ & $272$ & $3.53$ & $0.72$ & $0.81$ & $5.8\times 10^{5}$ & $150$ & $0.059$ & $18.6$ \\
MH-2 & $500$ & $256^3$ & $272$ & $3.53$ & $0.72$ & $0.9$ & $3.0\times 10^{5}$ & $110$ & $0.055$ & $19.5$ \\
MH-3 & $250$ & $128^3$ & $272$ & $3.53$ & $0.72$ & $1.2$ & $2.3\times 10^{5}$ & $94$ & $0.073$ & $20.9$ \\
MH-4 & $500$ & $256^3$ & $272$ & $3.53$ & $0.72$ & $1.1$ & $3.1\times 10^{5}$ & $97$ & $0.044$ & $22.6$ \\
MH-5 & $500$  & $256^3$ & $272$ & $3.53$ & $0.72$ & $1.3$ & $1.8\times 10^{5}$ & $58$ & $0.038 $ & $31.7$ \\ \hline
\multicolumn{11}{l}{} \\
\multicolumn{11}{l}{\parbox{16.5cm}{The comoving box sizes, particle numbers, initial DM masses, refined DM masses, gas masses, and normalizations used in the simulations, as well as the viral masses, virial radii, spin parameters, and collapse redshifts of the first minihalos that form. The halo properties agree well with the results of previous studies \citep[e.g.,][]{mba01,yoshida03a,gao07, on07}.}}\\
\end{tabular}
\end{center}
\end{table*}

In contrast, studies of present-day star formation have generally found fragmentation to occur shortly after the formation of the first protostar \citep{kb00,bbb03,krumholz09,peters10}. Following up on this result, sink particles were recently used in studies of the fragmentation of primordial gas in minihalos \citep{cgk08,sgb10,clark11a}. They found that the metal-free gas clouds fragment strongly, with the details of the process depending on the degree of turbulence in the halo. Focusing on the dynamical evolution of the high-density gas in the central regions of a minihalo, \citet{clark11b} demonstrated that the protostellar disks around primordial stars accrete from the infalling envelope faster than they can transfer their mass onto the central object. As a result, they rapidly become gravitationally unstable and fragment to build-up binary or higher-order multiple stellar systems. These results challenge the idea that the first stars formed in isolation and give rise to a number of new questions: What is the mass spectrum of Pop~III stars in groups? Is it different from the previously preferred single mode of star formation? How does it depend on the dynamical characteristics of the host halo? How does it influence subsequent cosmic evolution?

To address these issues, we here report on a series of hydrodynamical simulations performed with the quasi-Lagrangian moving mesh code {\small AREPO} \citep{springel10a}. Its high accuracy, flexibility, and efficiency allow us to follow the evolution of the gas in five statistically independent minihalos from fully cosmological initial conditions down to the optically thick regime where primordial protostars are born. We use a sink-particle approach that is similar to previously employed techniques to capture the subsequent accretion phase \citep[e.g.,][]{bbp95,kmk04, jappsen05, federrath10}, which allows us to reach well beyond the formation of the first protostar, and circumvent the limitations posed in previous high-resolution {\it ab initio} calculations of primordial star formation \citep{yoh08}. The five realizations give us an indication of how common fragmentation is, and of the shape of the mass function of Pop~III protostars during the early stages of accretion.

The structure of our work is as follows: In Section~2, we describe the numerical setup and physical ingredients of the simulations. In Section~3, we present the results of the simulations, followed by a discussion of radiation feedback, a resolution study, and the caveats of the sink particle treatment. Finally, in Section~4 we summarize our results and draw conclusions. All distances quoted in this paper are in proper units, unless noted otherwise.

\begin{figure*}
\begin{center}
\resizebox{12cm}{13.5cm}
{\unitlength1cm
\begin{picture}(12,13.5)
\put(0,7.5){\includegraphics[width=6cm,height=6cm]{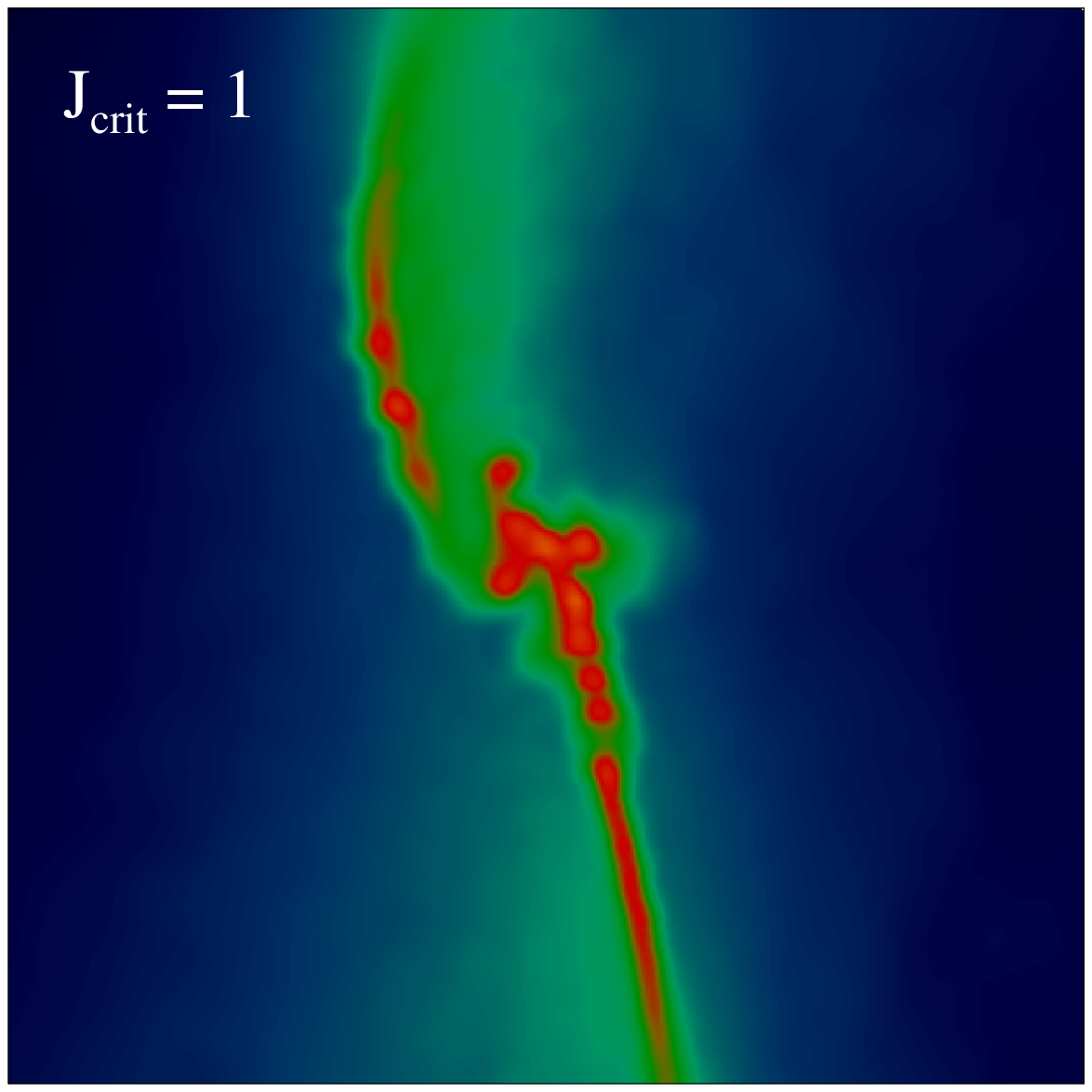}}
\put(6,7.5){\includegraphics[width=6cm,height=6cm]{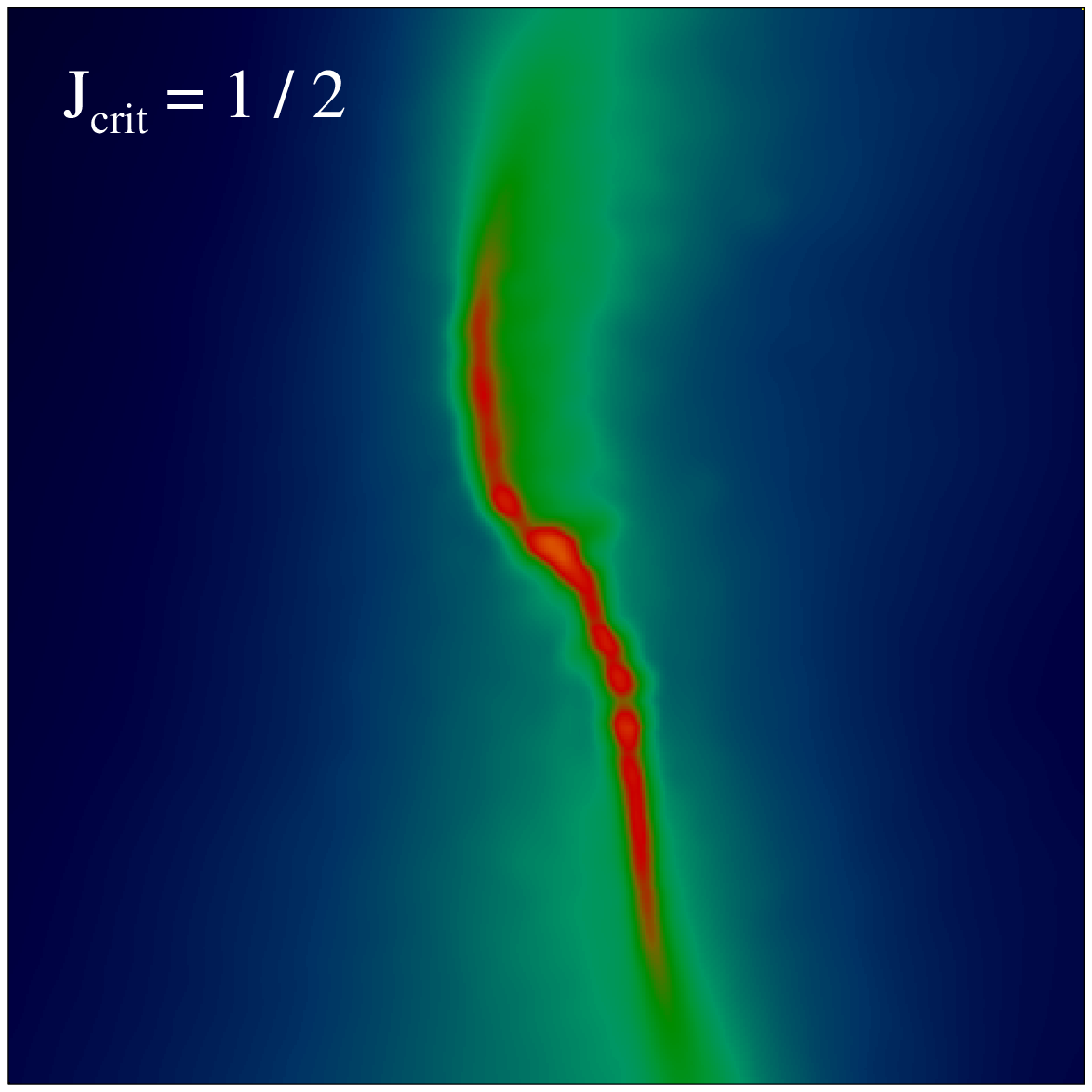}}
\put(0,1.5){\includegraphics[width=6cm,height=6cm]{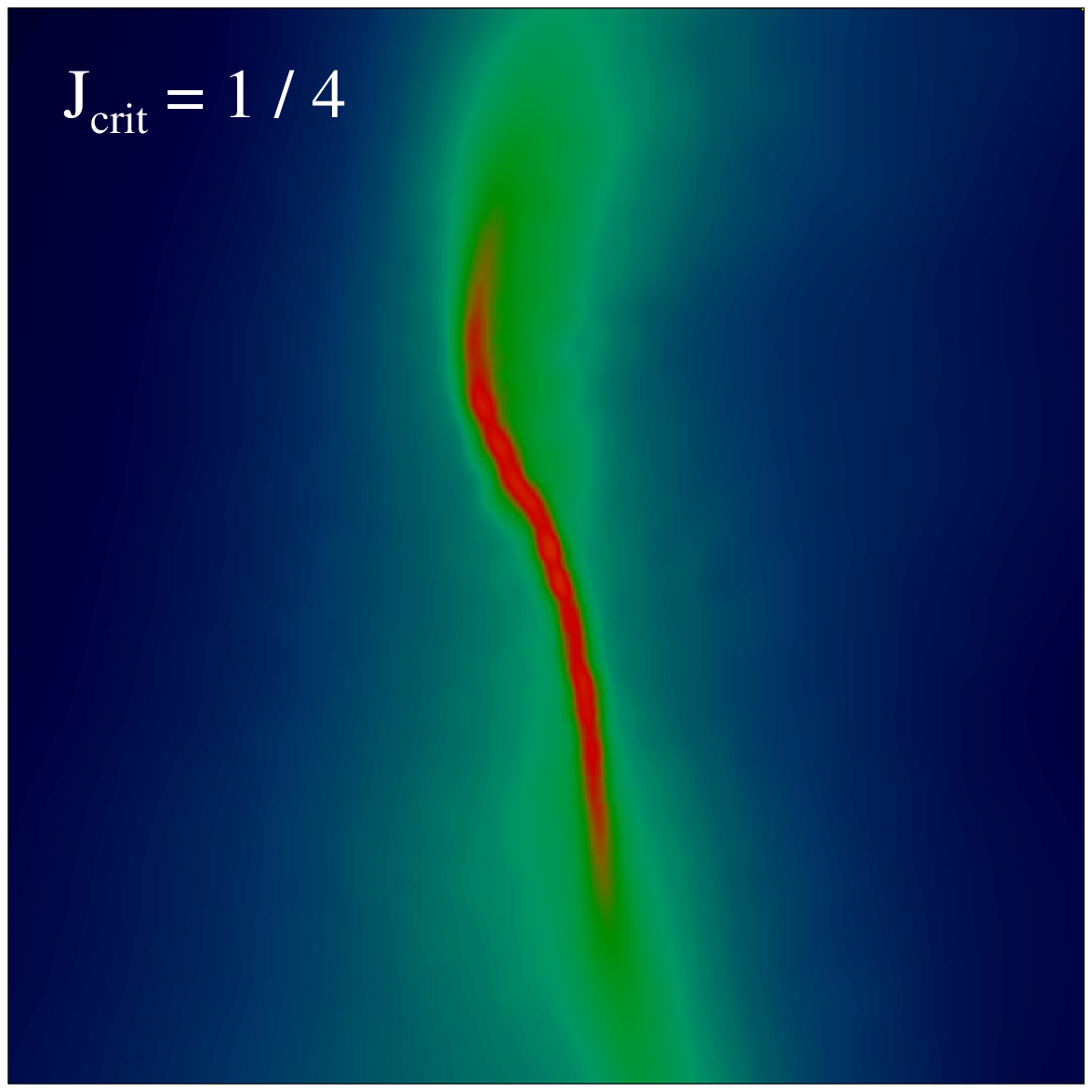}}
\put(6,1.5){\includegraphics[width=6cm,height=6cm]{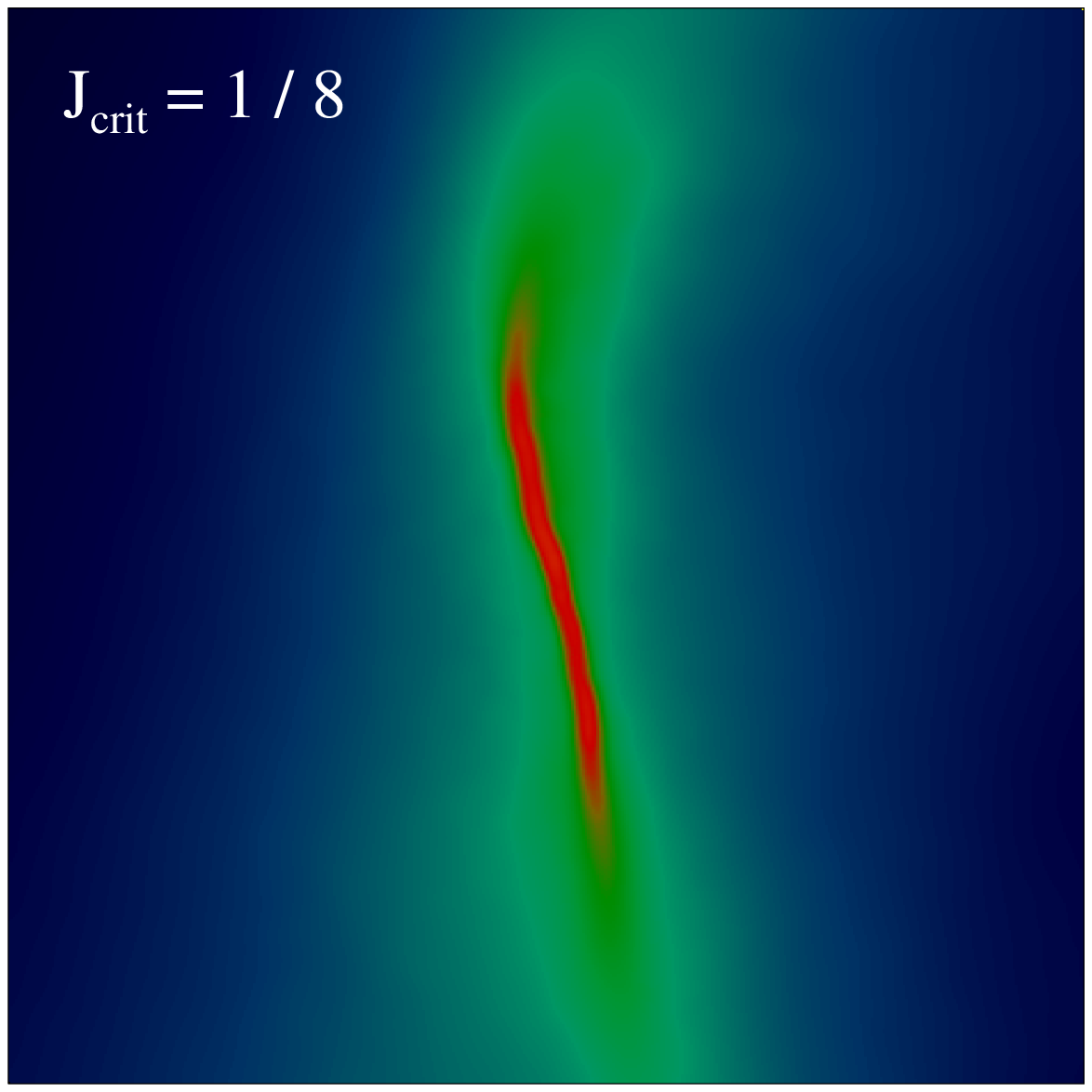}}
\put(0,0){\includegraphics[width=12cm,height=1.5cm]{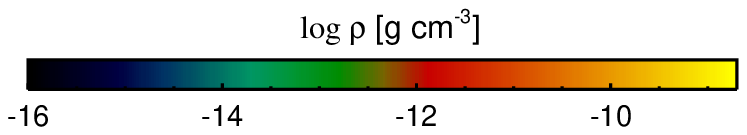}}
\end{picture}}
\caption{A test of the classical \citet{truelove98} criterion in {\scriptsize AREPO} for the \citet{bb79} isothermal collapse problem. The individual panels compare simulations with on-the-fly mesh refinement with $1$, $2$, $4$, and $8$ cells per Jeans length, respectively. We show the density-squared weighted mass density projected along the line of sight in a box with $200\,{\rm AU}$ on a side, centered on one of the two main clumps. Similar to the result found in AMR codes, the amount of artificial fragmentation increases dramatically once the local Jeans length is resolved by less than four cells.}
\end{center}
\end{figure*}

\begin{figure*}
\begin{center}
\resizebox{13.5cm}{16cm}
{\unitlength1cm
\begin{picture}(13.5,16)
\put(0,10.5){\includegraphics[width=4.5cm,height=4.5cm]{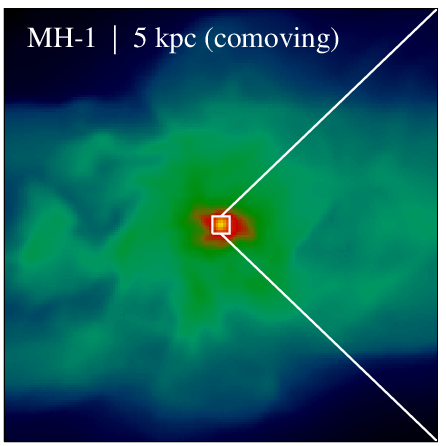}}
\put(4.5,10.5){\includegraphics[width=4.5cm,height=4.5cm]{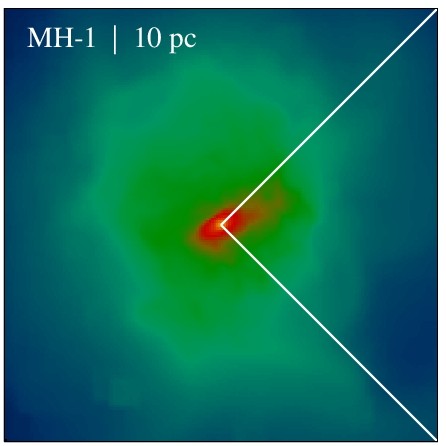}}
\put(9,10.5){\includegraphics[width=4.5cm,height=4.5cm]{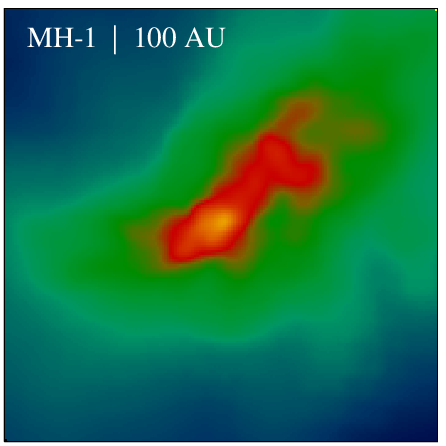}}
\put(0,6){\includegraphics[width=4.5cm,height=4.5cm]{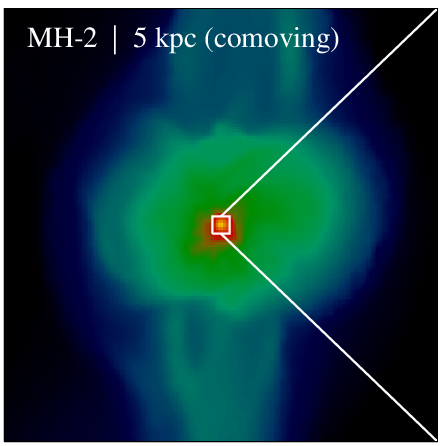}}
\put(4.5,6){\includegraphics[width=4.5cm,height=4.5cm]{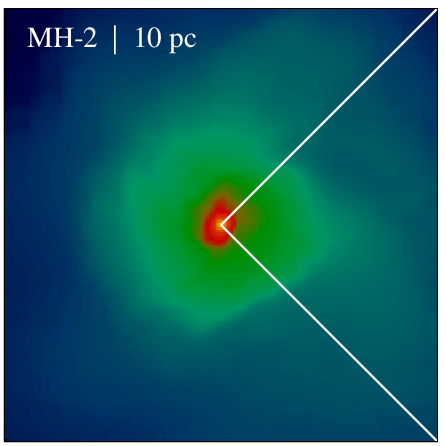}}
\put(9,6){\includegraphics[width=4.5cm,height=4.5cm]{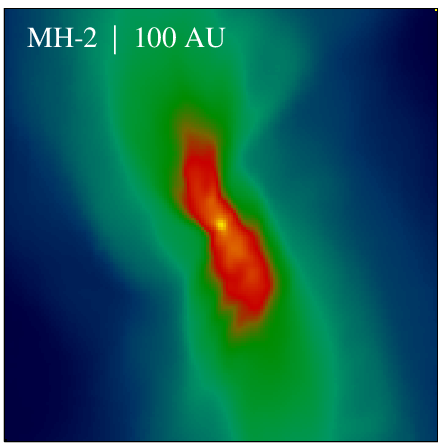}}
\put(0,1.5){\includegraphics[width=4.5cm,height=4.5cm]{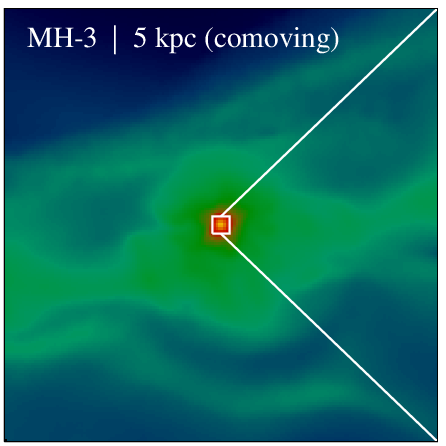}}
\put(4.5,1.5){\includegraphics[width=4.5cm,height=4.5cm]{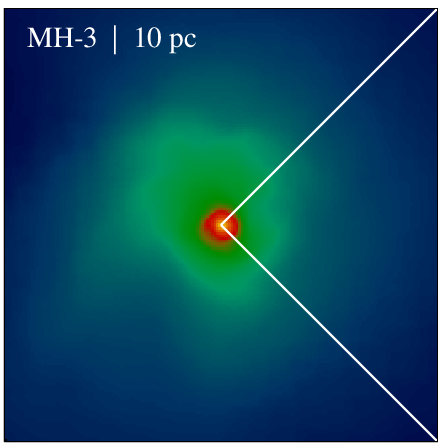}}
\put(9,1.5){\includegraphics[width=4.5cm,height=4.5cm]{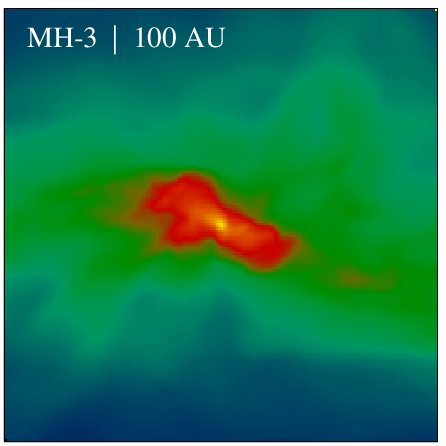}}
\put(0,0){\includegraphics[width=4.5cm,height=1.5cm]{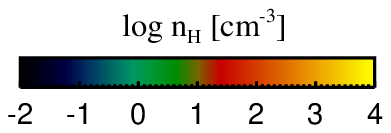}}
\put(4.5,0){\includegraphics[width=4.5cm,height=1.5cm]{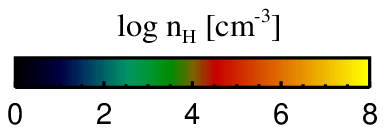}}
\put(9,0){\includegraphics[width=4.5cm,height=1.5cm]{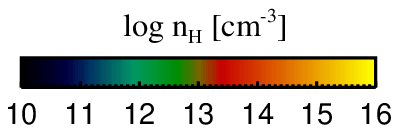}}
\end{picture}}
\caption{A sequential zoom-in on the gas in each of the five minihalos. The panels show the density-squared weighted number density of hydrogen nuclei projected along the line of sight. The gas virializes on a scale of $\simeq 5\,{\rm kpc}$ (comoving), followed by the runaway collapse of the central $\simeq 1\,{\rm pc}$, where the gas becomes self-gravitating and decouples from the dark matter. In the final stages of the collapse, a fully molecular core on a scale of a few hundred AU forms. The visible turbulence induced by the virialization of the dark matter halo survives down to the smallest scales and later influences the fragmentation of the gas.}
\end{center}
\end{figure*}

\begin{figure*}
\begin{center}
\resizebox{13.5cm}{10.5cm}
{\unitlength1cm
\begin{picture}(13.5,10.5)
\put(0,6){\includegraphics[width=4.5cm,height=4.5cm]{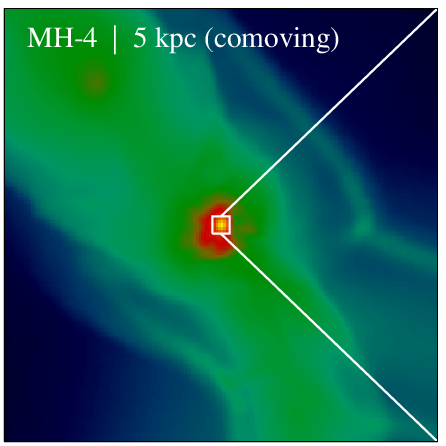}}
\put(4.5,6){\includegraphics[width=4.5cm,height=4.5cm]{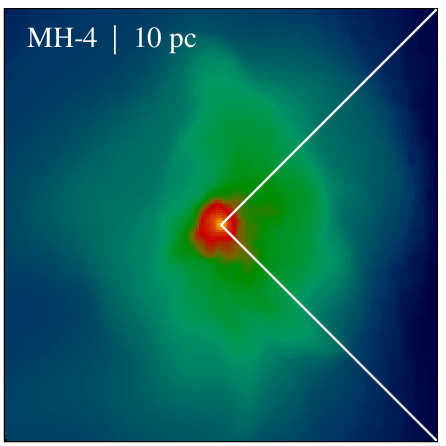}}
\put(9,6){\includegraphics[width=4.5cm,height=4.5cm]{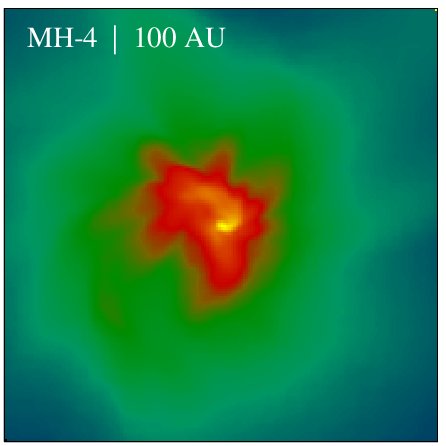}}
\put(0,1.5){\includegraphics[width=4.5cm,height=4.5cm]{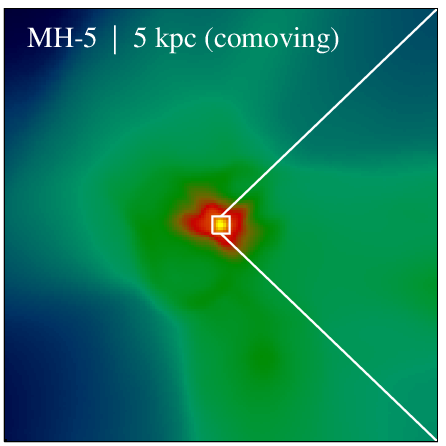}}
\put(4.5,1.5){\includegraphics[width=4.5cm,height=4.5cm]{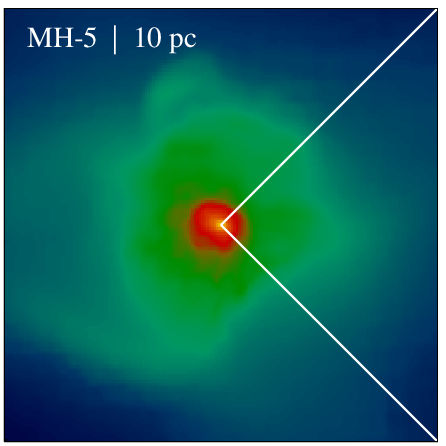}}
\put(9,1.5){\includegraphics[width=4.5cm,height=4.5cm]{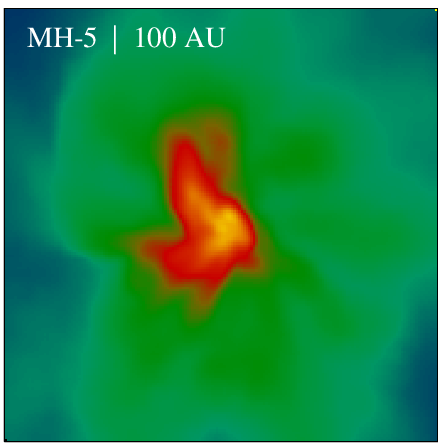}}
\put(0,0){\includegraphics[width=4.5cm,height=1.5cm]{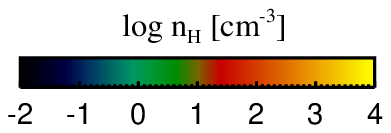}}
\put(4.5,0){\includegraphics[width=4.5cm,height=1.5cm]{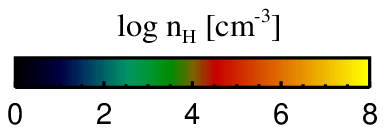}}
\put(9,0){\includegraphics[width=4.5cm,height=1.5cm]{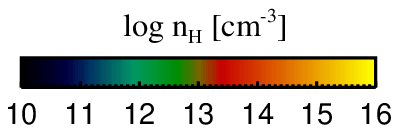}}
\end{picture}}
\caption{See Figure~2 for caption.}
\end{center}
\end{figure*}

\begin{figure*}
\begin{center}
\includegraphics[width=12cm]{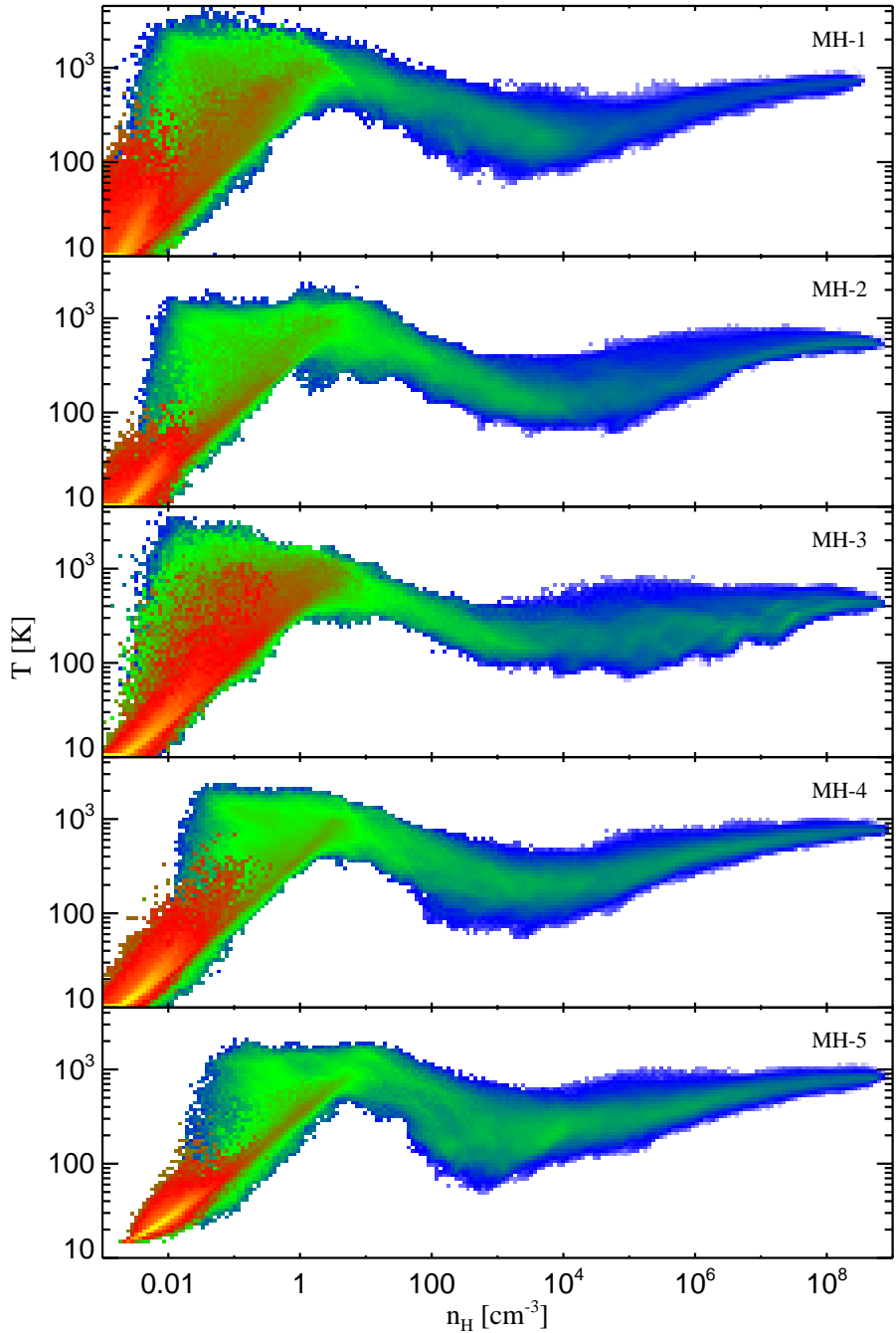}
\caption{Temperature versus number density of hydrogen nuclei in the fully cosmological simulations. The mass in each bin is color-coded from blue (minimum) to yellow (maximum). The evolution of the gas is similar to the results of previous studies, with the exception that HD cooling becomes important in simulations MH-2 and MH-3 and leads to a prolonged cooling phase at $n_{\rm H}\ga 10^4\,{\rm cm}^{-3}$.}
\end{center}
\end{figure*}

\begin{figure*}
\begin{center}
\includegraphics[width=12cm]{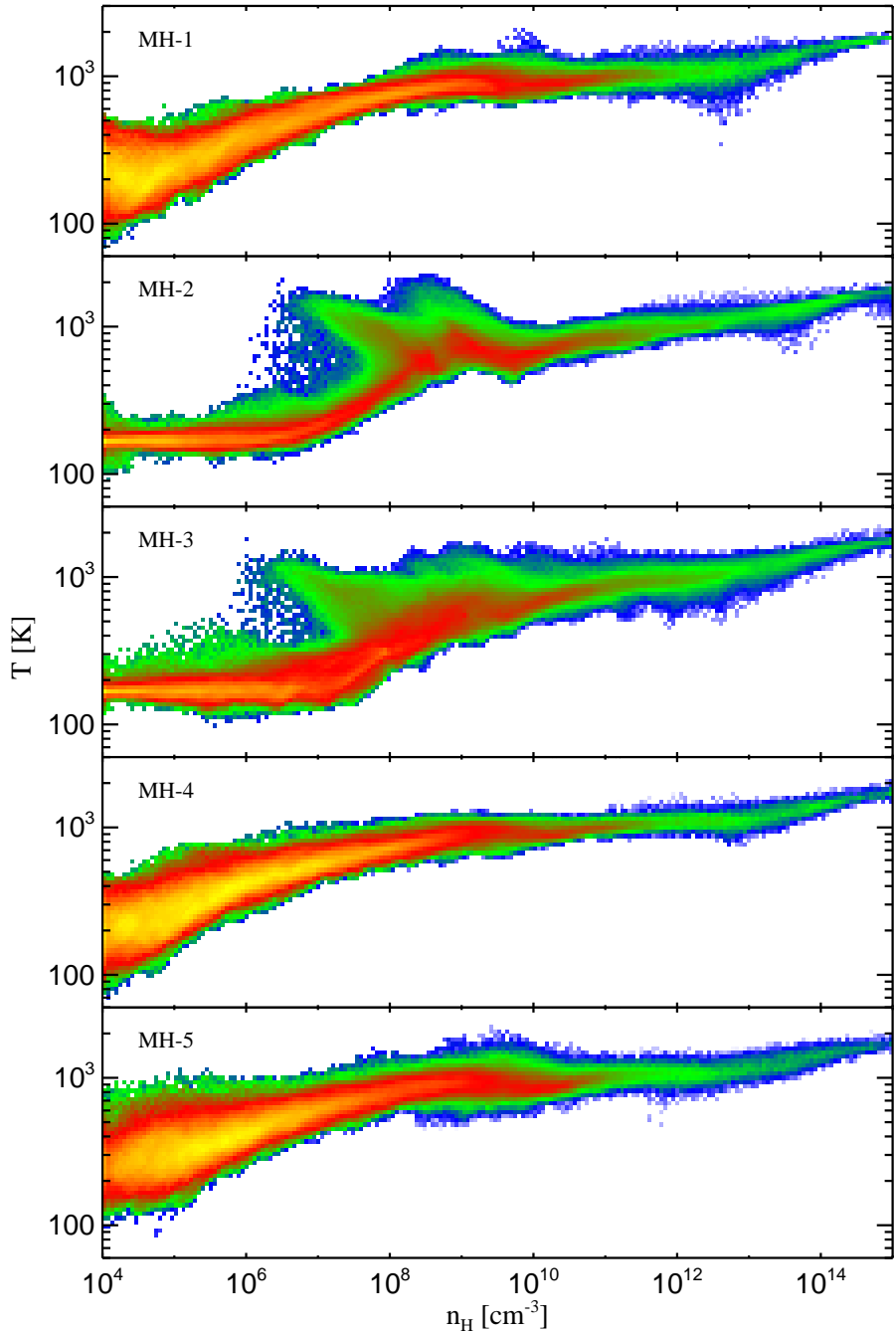}
\caption{Same as Figure~4, but for the follow-up high resolution simulations. In simulations MH-2 and MH-3, the gas heats up more violently after the prolonged cooling phase due to HD cooling, which creates shocks that are visible as `fingers' of hot, underdense gas. At densities greater than $\sim 10^{16}\,{\rm cm}^{-3}$, the gas becomes optically thick to cooling radiation and shortly thereafter forms a protostar.}
\end{center}
\end{figure*}

\begin{figure*}
\begin{center}
\includegraphics[width=14cm]{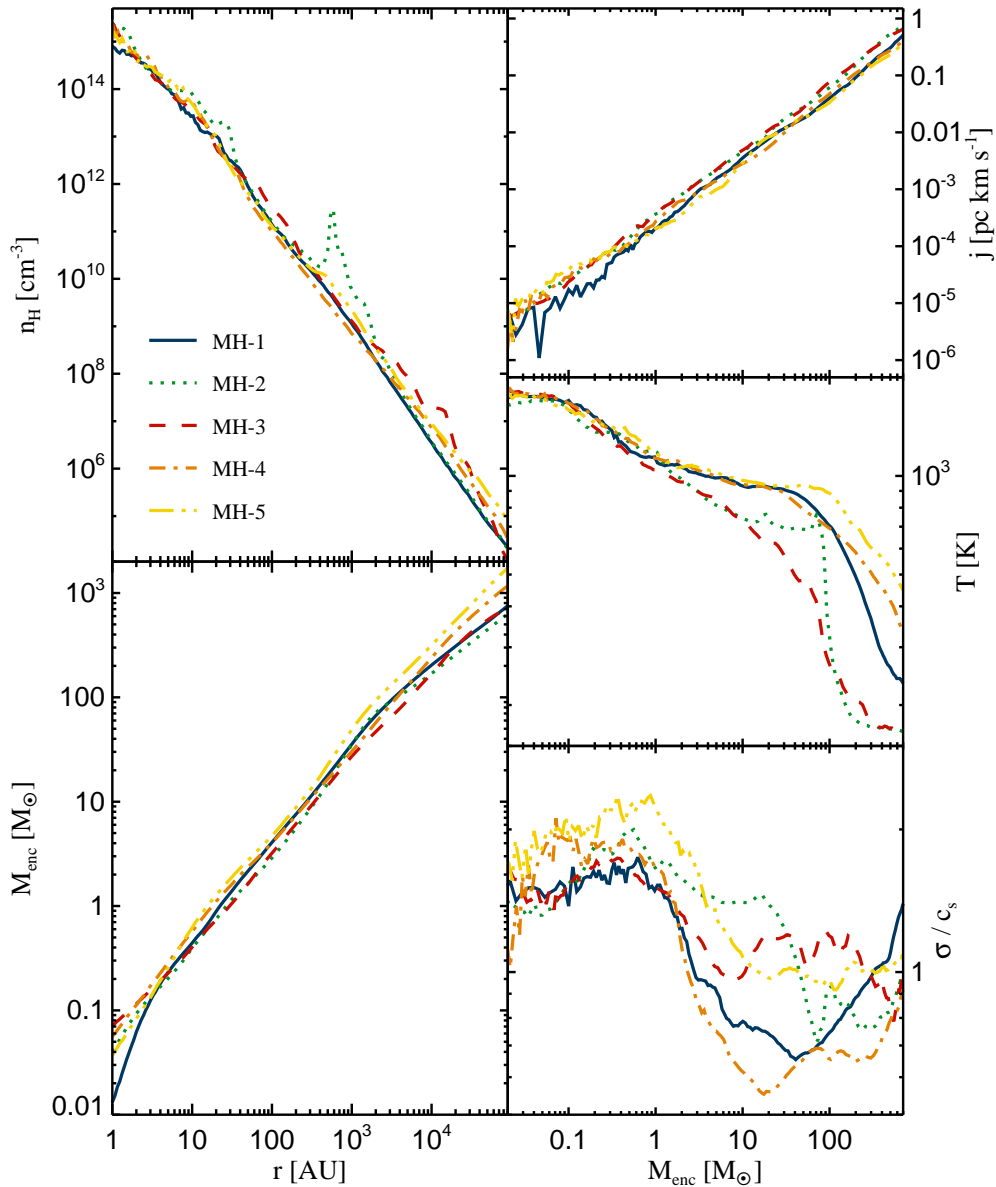}
\caption{The panels on the left show the spherically averaged number density of hydrogen nuclei and enclosed gas mass as a function of radius just before the formation of the first protostar. The panels on the right show the spherically averaged specific angular momentum, temperature, and velocity dispersion in units of the sound speed as a function of enclosed gas mass. The density, enclosed gas mass, and angular momentum profiles are very similar overall, while the thermal evolution of the gas displays some scatter due to the activation of HD cooling in simulations MH-2 and MH-3. In these two minihalos, the gas heats up later but more violently after becoming gravitationally unstable. The velocity dispersion shows no convincing correlation with the thermal history of the gas and is always close to Mach numbers $M\simeq 1$, indicating transsonic turbulence. An interesting feature is the simultaneous collapse of a second gas cloud in simulation MH-2, which is visible as a density peak at about $1000\,{\rm AU}$ from the primary cloud. Similar behavior was found in another recent study \citep{tao09}.}
\end{center}
\end{figure*}

\begin{figure*}
\begin{center}
\resizebox{16cm}{21.5cm}
{\unitlength1cm
\begin{picture}(16,21.5)
\put(0,17.5){\includegraphics[width=4cm,height=4cm]{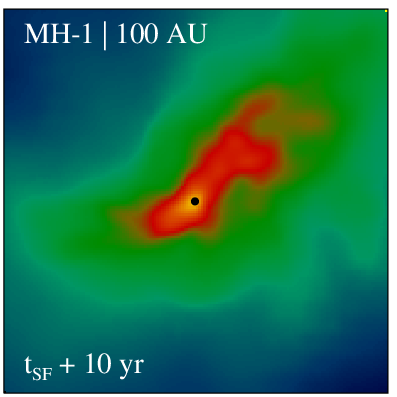}}
\put(4,17.5){\includegraphics[width=4cm,height=4cm]{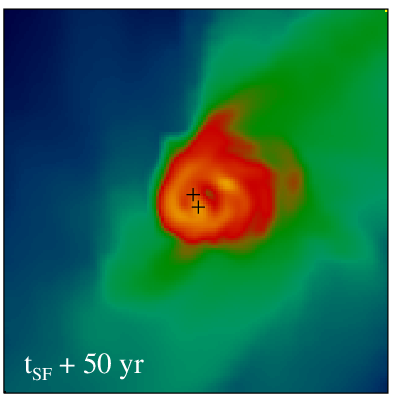}}
\put(8,17.5){\includegraphics[width=4cm,height=4cm]{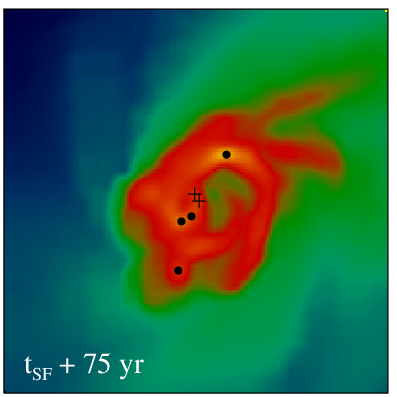}}
\put(12,17.5){\includegraphics[width=4cm,height=4cm]{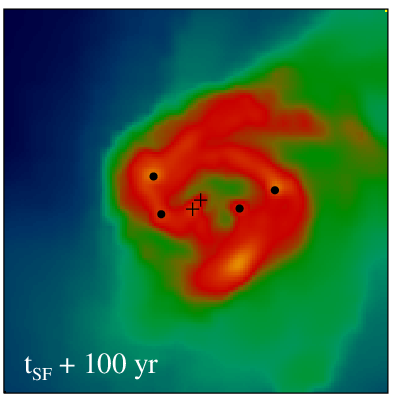}}
\put(0,13.5){\includegraphics[width=4cm,height=4cm]{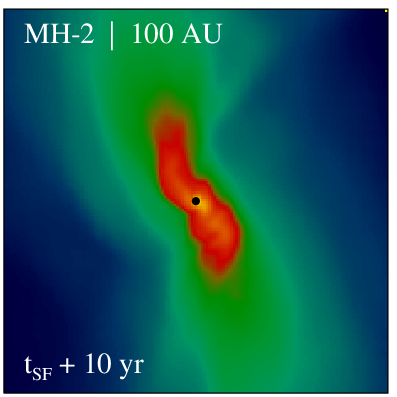}}
\put(4,13.5){\includegraphics[width=4cm,height=4cm]{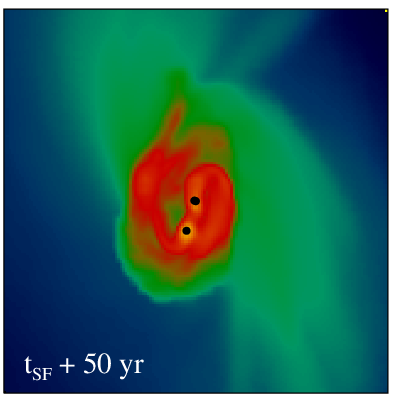}}
\put(8,13.5){\includegraphics[width=4cm,height=4cm]{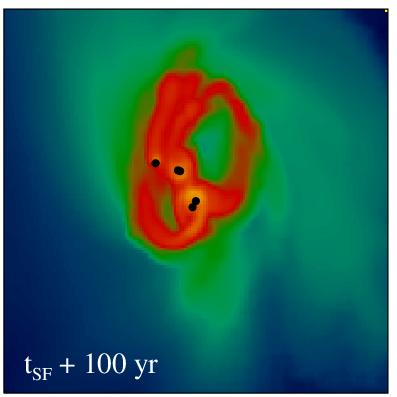}}
\put(12,13.5){\includegraphics[width=4cm,height=4cm]{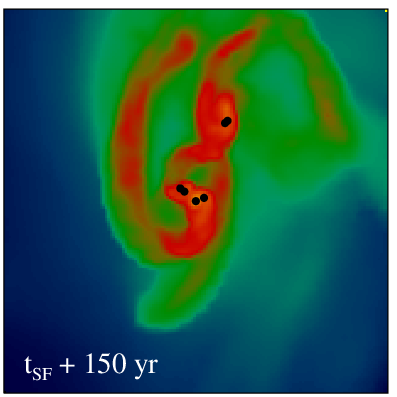}}
\put(0,9.5){\includegraphics[width=4cm,height=4cm]{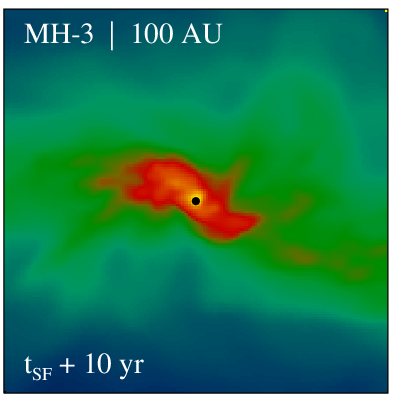}}
\put(4,9.5){\includegraphics[width=4cm,height=4cm]{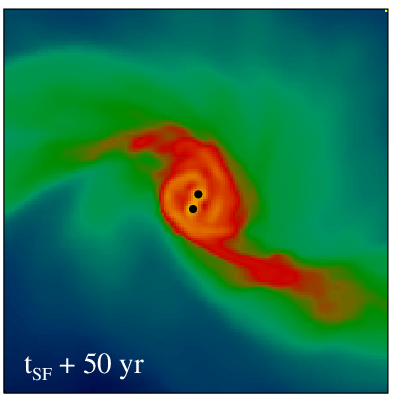}}
\put(8,9.5){\includegraphics[width=4cm,height=4cm]{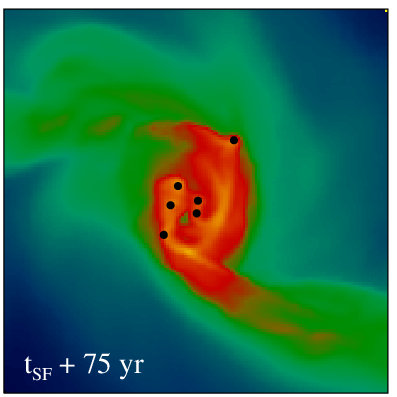}}
\put(12,9.5){\includegraphics[width=4cm,height=4cm]{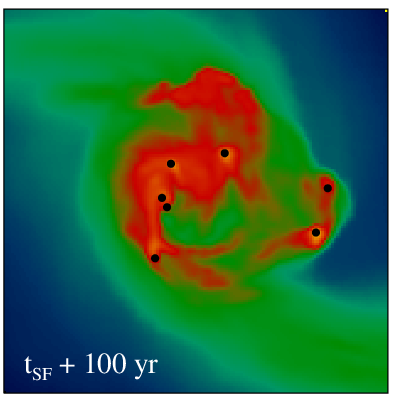}}
\put(0,5.5){\includegraphics[width=4cm,height=4cm]{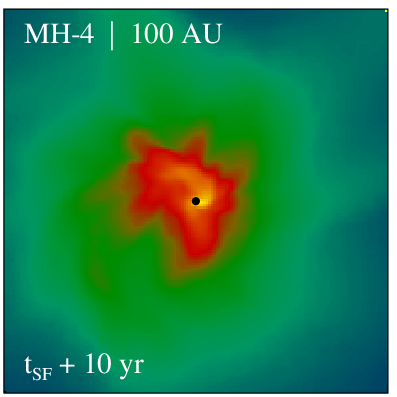}}
\put(4,5.5){\includegraphics[width=4cm,height=4cm]{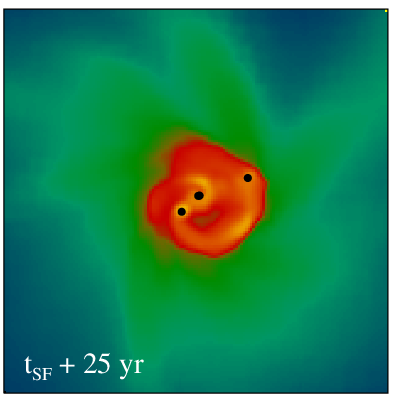}}
\put(8,5.5){\includegraphics[width=4cm,height=4cm]{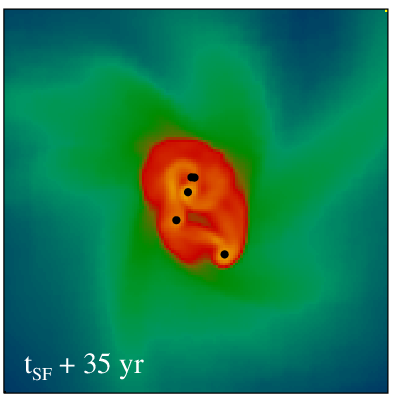}}
\put(12,5.5){\includegraphics[width=4cm,height=4cm]{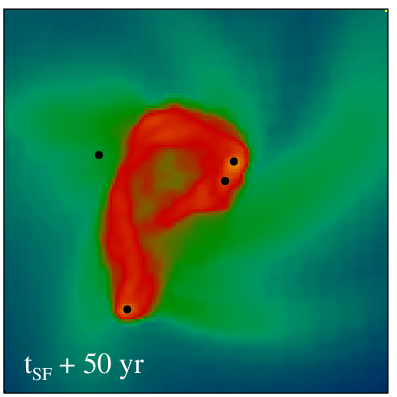}}
\put(0,1.5){\includegraphics[width=4cm,height=4cm]{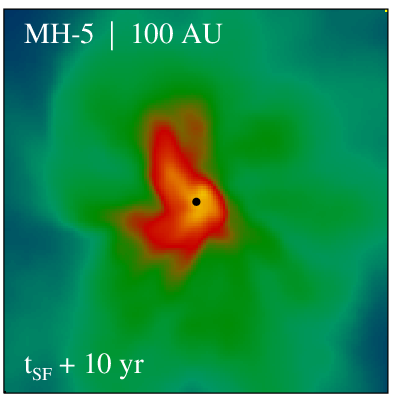}}
\put(4,1.5){\includegraphics[width=4cm,height=4cm]{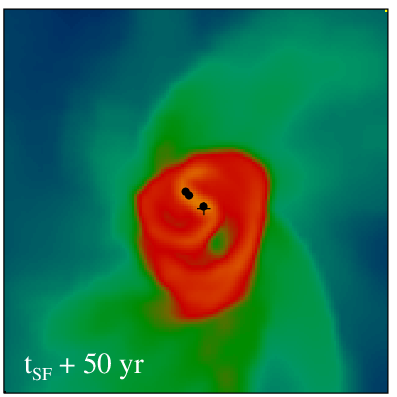}}
\put(8,1.5){\includegraphics[width=4cm,height=4cm]{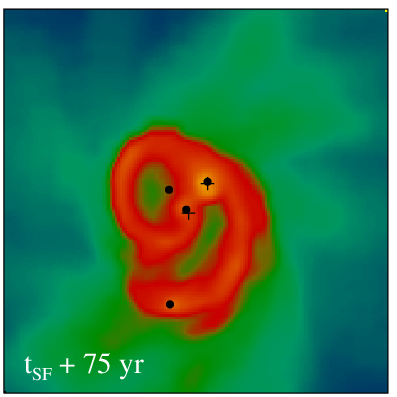}}
\put(12,1.5){\includegraphics[width=4cm,height=4cm]{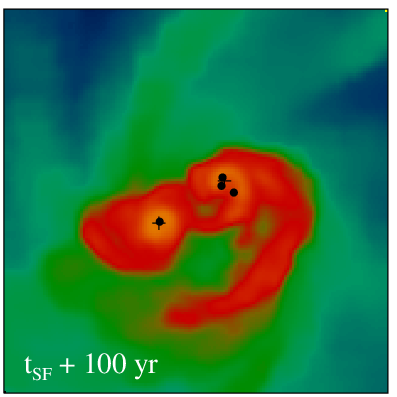}}
\put(0,0){\includegraphics[width=16cm,height=1.5cm]{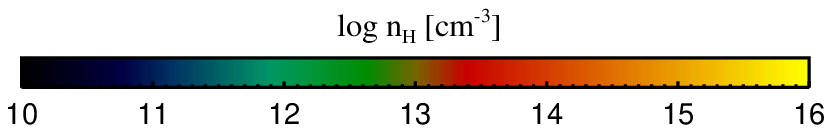}}
\end{picture}}
\caption{The formation of a protostellar cluster at the center of the minihalos using standard sink particles. The panels show the density-squared weighted number density of hydrogen nuclei projected along the line of sight. Black dots and crosses denote protostars with masses below and above $1\,{\rm M}_\odot$, respectively. The process of initial disk formation and fragmentation is remarkably similar in all minihalos \citep[see also][]{clark11b}, after which N-body effects become important and lead to relatively unique configurations. For example, in simulation MH-4 dynamical interactions have led to the ejection of a low-mass protostar after only $\simeq 50\,{\rm yr}$. This occurs significantly later in the other four minihalos.}
\end{center}
\end{figure*}


\begin{figure*}
\begin{center}
\resizebox{13.5cm}{10.5cm}
{\unitlength1cm
\begin{picture}(13.5,10.5)
\put(0,6){\includegraphics[width=4.5cm,height=4.5cm]{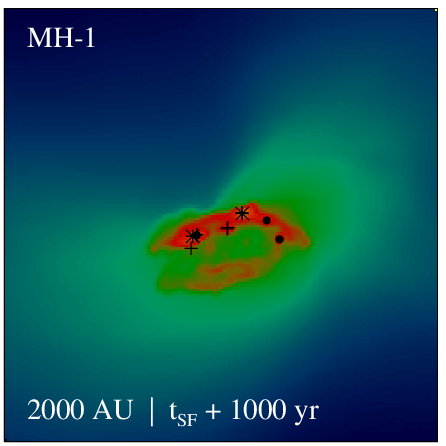}}
\put(4.5,6){\includegraphics[width=4.5cm,height=4.5cm]{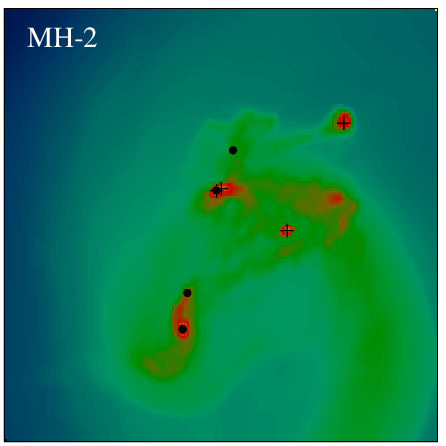}}
\put(9,6){\includegraphics[width=4.5cm,height=4.5cm]{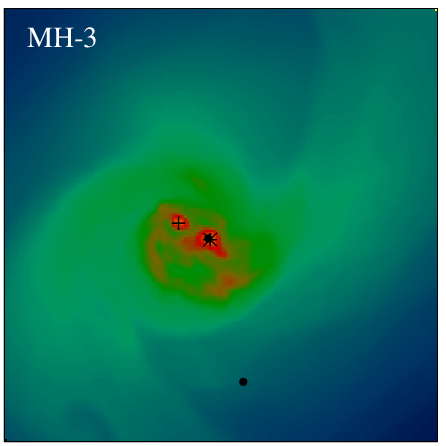}}
\put(2.25,1.5){\includegraphics[width=4.5cm,height=4.5cm]{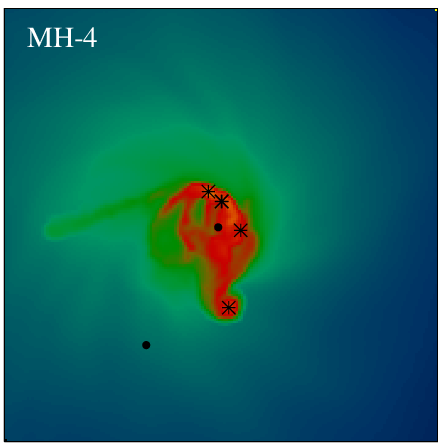}}
\put(6.75,1.5){\includegraphics[width=4.5cm,height=4.5cm]{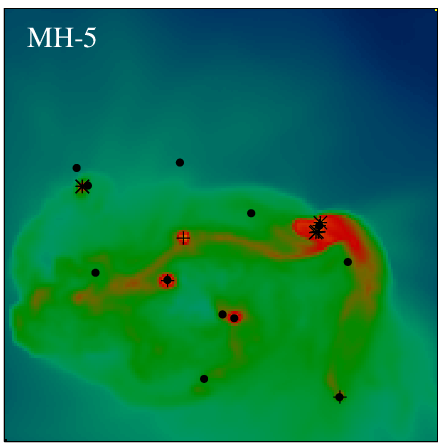}}
\put(0,0){\includegraphics[width=13.5cm,height=1.5cm]{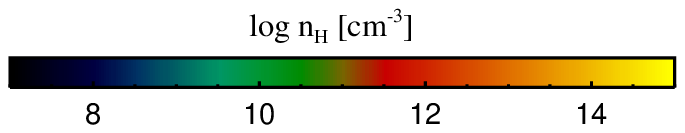}}
\end{picture}}
\caption{The central $2000\,{\rm AU}$ after $1000\,{\rm yr}$ of continued fragmentation and accretion. Black dots, crosses and stars denote protostars with masses below $1\,{\rm M}_\odot$, between $1\,{\rm M}_\odot$ and $3\,{\rm M}_\odot$, and above $3\,{\rm M}_\odot$. A relatively rich protostellar cluster with a range of masses has survived in each case. In a few minihalos, low-mass protostars have been ejected out of the central gas cloud, such that they are no longer visible here. In simulation MH-2, two independent clumps have collapsed almost simultaneously and formed their own clusters before eventually merging (see also Figure~6).}
\end{center}
\end{figure*}

\begin{figure*}
\begin{center}
\includegraphics[width=14cm]{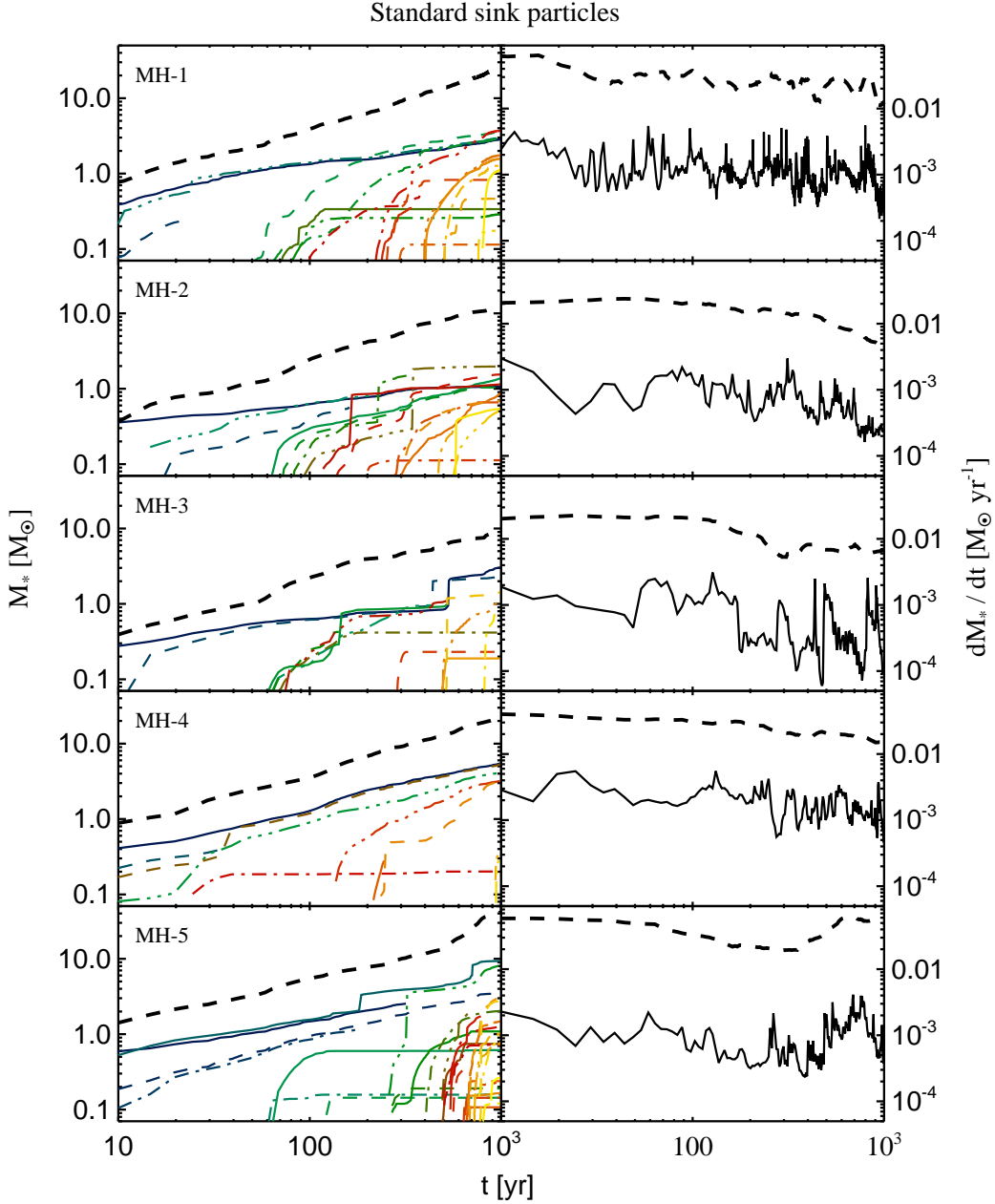}
\caption{The mass accretion histories of the protostars (left panels), and the average accretion rates of the entire ensemble of protostars (right panels) using standard sink particles. In the left panels, each line denotes the evolution of an individual protostar. The thick dashed lines in both panels denote the cumulative values. Mergers are indicated by instantaneous jumps in mass. We find that star formation is not restricted to a single burst, but occurs continually for the entire simulated timespan. In every minihalo, between $5$ and $15$ protostars with masses ranging from $0.1$ to nearly $10\,{\rm M}_\odot$ are formed. The total accretion rates are nearly constant over time at a few $0.01\,{\rm M}_{\odot}\,{\rm yr}^{-1}$.}
\end{center}
\end{figure*}

\begin{figure*}
\begin{center}
\includegraphics[width=14cm]{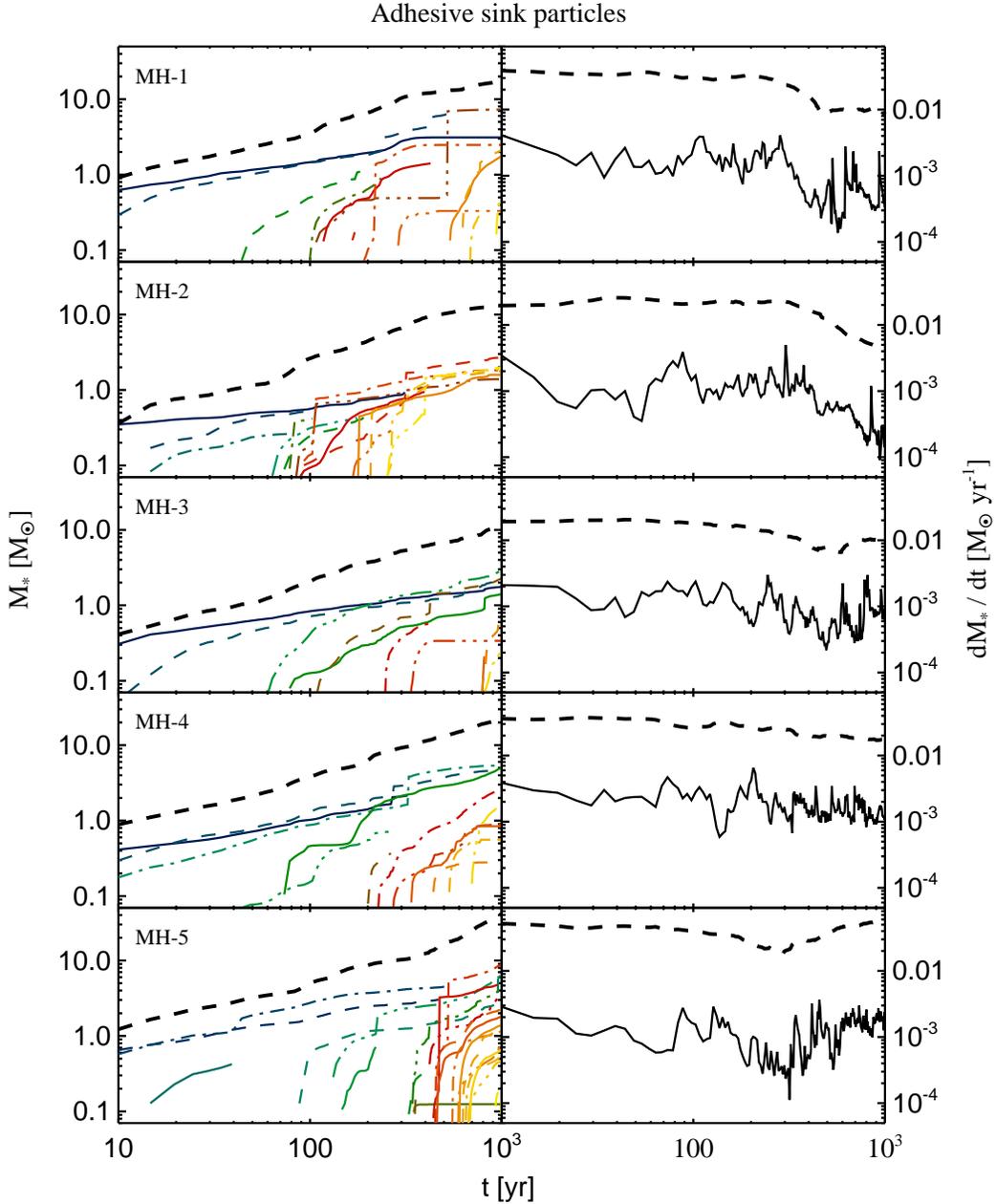}
\caption{Same as Figure~9, but using adhesive sink particles. In this case fewer protostars with systematically higher masses are formed, although the total amount of gas within protostars is nearly identical.}
\end{center}
\end{figure*}

\begin{figure*}
\begin{center}
\includegraphics[width=14cm]{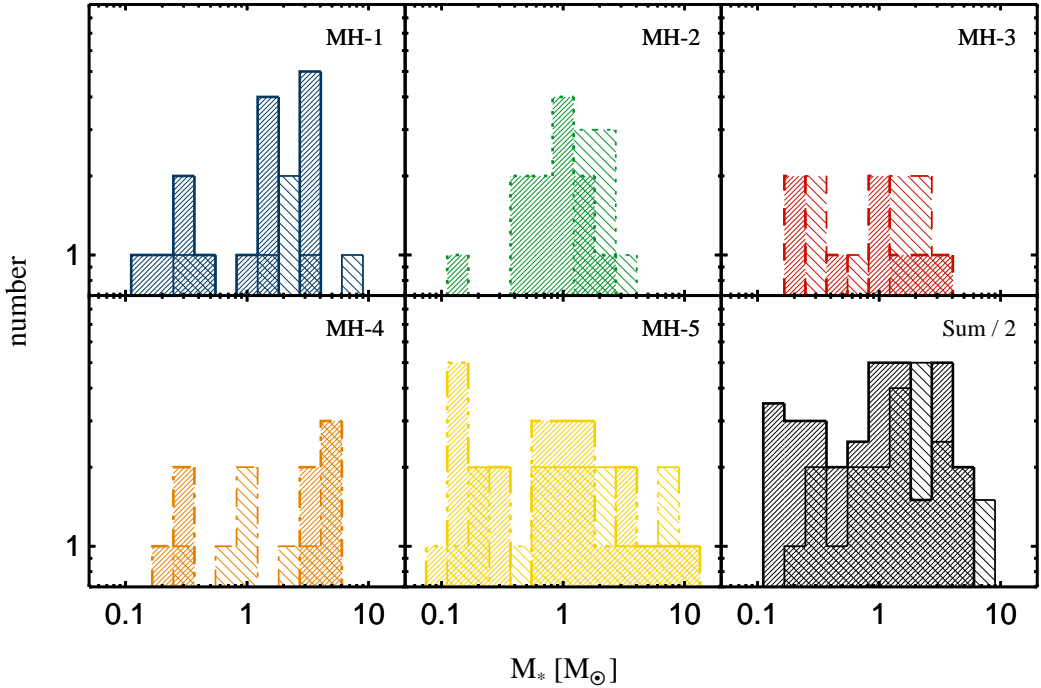}
\caption{The protostellar mass function after $1000$ years of continued fragmentation and accretion. The dark and light shadings distinguish the mass functions obtained for standard and adhesive sink particles, respectively. Despite very aggressive merging, a small cluster of protostars with a range of masses is formed even in the latter case. In the bottom right panel, we also show the cumulative mass functions obtained by summing up the contributions from the individual minihalos, and renormalized for better visibility. The resulting distribution is relatively flat between $\sim 0.1$ and $\sim 10\,{\rm M}_\odot$, indicating that most of the mass is locked up in high-mass protostars.}
\end{center}
\end{figure*}

\begin{figure*}
\begin{center}
\includegraphics[width=14cm]{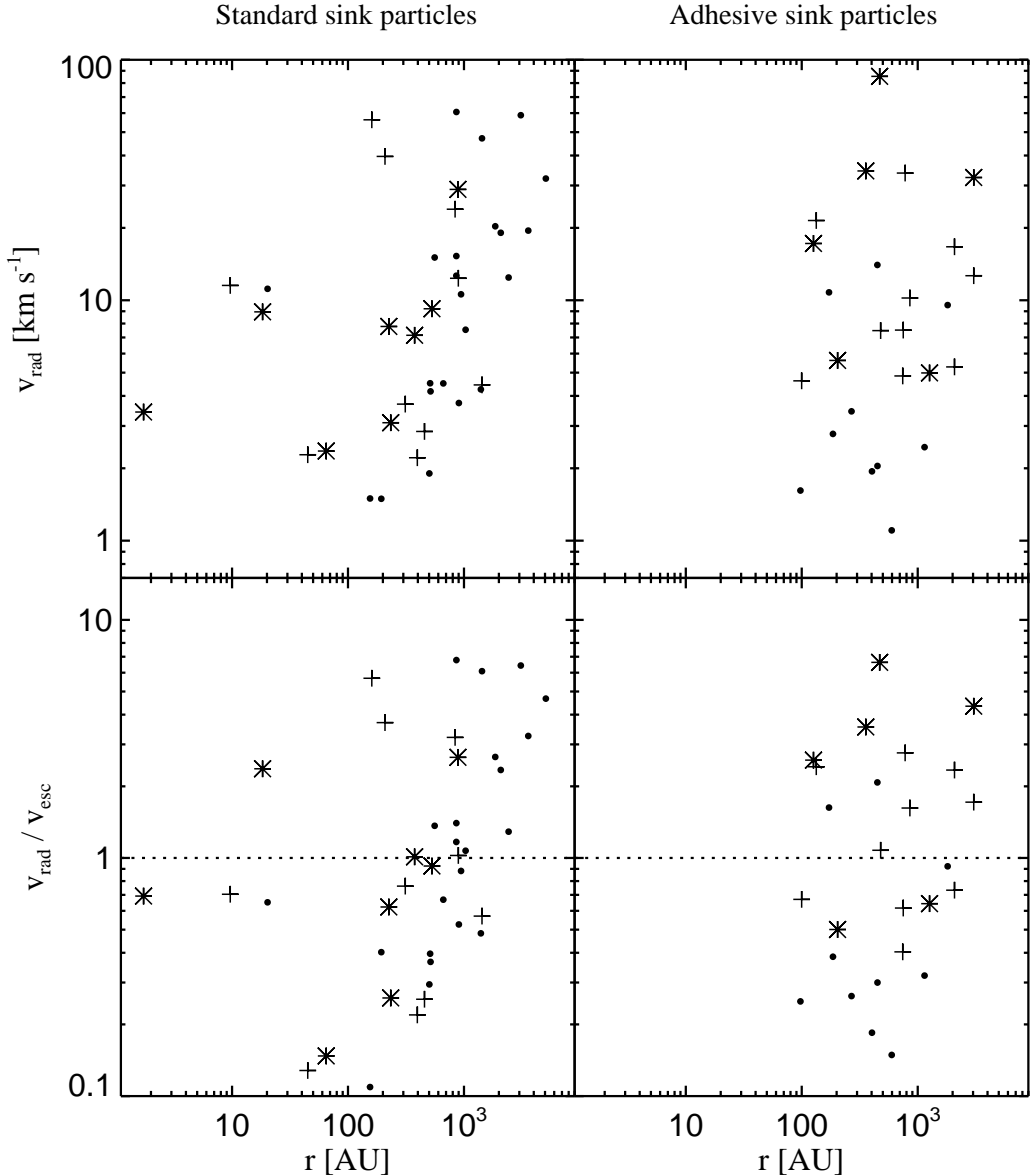}
\caption{The radial velocity and ratio of the radial velocity to the escape velocity of all protostars as a function of distance to the center of mass, shown for standard (left panels) and adhesive (right panels) sink particles. The critical ratio of unity, where protostars are assumed to escape from the central gas cloud, is denoted by the dashed line. The escape velocity is determined by using the total mass enclosed within the current distance of each protostar from the center of mass. Black dots, crosses and stars denote protostars with masses below $1\,{\rm M}_\odot$, between $1\,{\rm M}_\odot$ and $3\,{\rm M}_\odot$, and above $3\,{\rm M}_\odot$. In our standard implementation of sink particles, a number of low-mass protostars obtain high radial velocities and escape from the central gas cloud. They stop accreting after they receive substantial radial velocities during close encounters with other protostars. For adhesive sink particles, this occurs significantly less often. In this case no protostars reside within the central $\simeq 100\,{\rm AU}$, since momentum conservation acts to move merged protostars to larger annuli.}
\end{center}
\end{figure*}

\begin{figure*}
\begin{center}
\resizebox{12cm}{7.5cm}
{\unitlength1cm
\begin{picture}(12,7.5)
\put(0.0,1.5){\includegraphics[width=6cm,height=6cm]{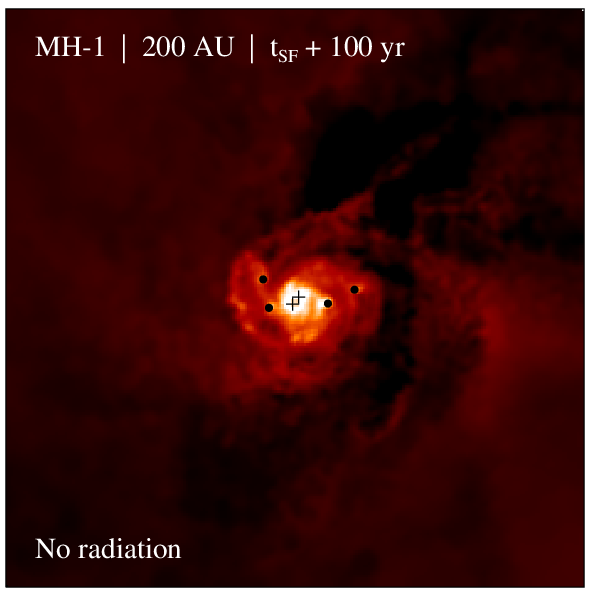}}
\put(6.0,1.5){\includegraphics[width=6cm,height=6cm]{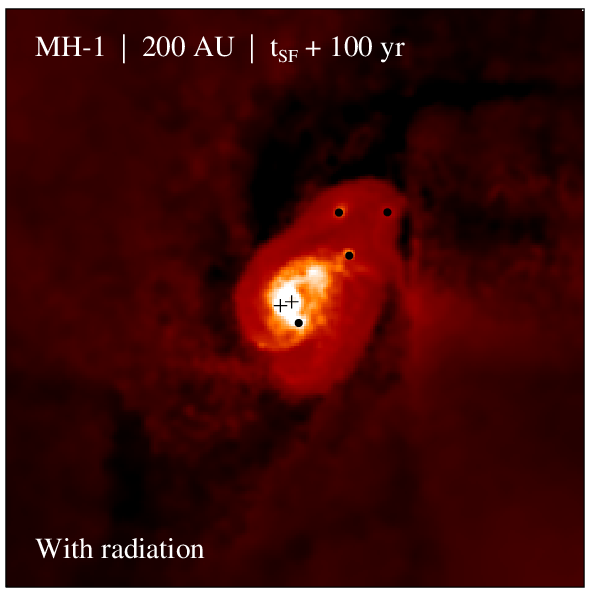}}
\put(0.0,0.0){\includegraphics[width=12cm,height=1.5cm]{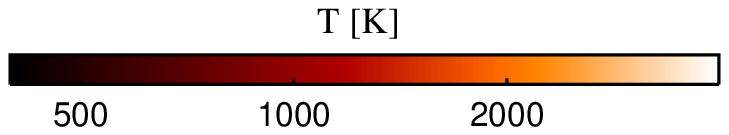}}
\end{picture}}
\caption{A comparison of the fragmentation of the gas in simulation MH-1 with and without infrared radiation emitted by accreting protostars. The panels show the density-squared weighted temperature projected along the line of sight. The radiation slightly heats and puffs up the disk, such that the fragments form at somewhat larger distances from the center. The qualitative nature of radiation feedback is therefore very similar to what studies of present-day star formation have shown \citep{krumholz09,peters10}. However, the effect is significantly reduced here due to the high temperature of the gas and its very efficient cooling by molecular hydrogen lines.}
\end{center}
\end{figure*}

\section{Numerical Methodology}

We here provide details about the set-up of the pure dark matter (DM) simulations, the employed resimulation technique, and the main simulation runs with the hydrodynamic moving mesh code {\small AREPO}. We also discuss the implementation of the on-the-fly mesh refinement technique, the sink-particle treatment, and the chemistry and cooling model.

\subsection{Dark Matter Simulations}

The DM-only simulations are initialized in cosmological boxes with $250$, $500$ and $1000\,{\rm kpc}$ (comoving) on a side, and employ $128^3$, $256^3$ and $512^3$ particles of mass $M_{\rm dm}\simeq 272\,{\rm M}_\odot$, so that minihalos with a characteristic mass of $5\times 10^5\,{\rm M}_\odot$ are resolved by $\simeq 2\times 10^3$ particles and a comoving gravitational softening length of $68\,{\rm pc}$. We use an initial fluctuation power spectrum for a $\Lambda$ cold dark matter ($\Lambda$CDM) cosmology at a starting redshift of $z=99$ with matter density $\Omega_m=1-\Omega_\Lambda=0.27$, baryon density $\Omega_b=0.046$, Hubble parameter $h=H_0/100\,{\rm km}\,{\rm s}^{-1}\,{\rm Mpc}^{-1}=0.71$ (where $H_0$ is the present Hubble expansion rate) and a spectral index $n_s=0.96$. These parameters are based on the five-year WMAP results \citep{komatsu09}. We use a range of values for the linearly extrapolated present-day normalization $\sigma_8$ (as given by Table~1), which allows us to mimic rare overdense regions of the universe that cannot be captured with the limited box sizes employed here.

All our DM simulations were evolved with the TreePM/SPH code {\small GADGET-3} \citep[last described in][]{springel05} until the first minihalo with a virial mass exceeding $5\times 10^5\,{\rm M}_\odot$ collapses, thereby providing a target region for our subsequent resimulations. In this study, we restrict ourselves to the analysis of five independent realizations of the minihalos identified in the large-scale DM runs, denoted MH-1 to MH-5. The most important parameters of the simulations and the properties of the minihalos are summarized in Table~1.

\subsection{Resimulation Technique}

Once the location of a minihalo in the parent simulation has been determined, we construct a high-resolution study of this object by first identifying all particles within the virial radius and a sufficiently large boundary region around it. These particles are then traced back to their initial conditions, yielding the Lagrangian region that formed the object. To construct new initial conditions, each particle in this region is replaced by $64$ DM particles and $64$ mesh-generating points which produce the space-filling cells that are used for our hydrodynamic calculations.

Cells and DM particles outside of the high-resolution region are replaced by ever higher mass particles with increasing distance from the target halo, such that the resolution coarsens significantly far away from the target halo, yet the gravitational tidal field that influences its formation is still followed with high accuracy. This step reduces the total number of particles and cells in the refined initial conditions to $\simeq 2\times 10^6$. The simulation box is then centered on the target halo and reinitialized at $z=99$ based on the original realization of the density field, but augmented with additional small-scale power in the high-resolution region that can now be represented. The initial DM particle and cell masses in the high-resolution region before any further run-time refinement are given by $M_{\rm dm, ref}=(1-\Omega_b/\Omega_m)M_{\rm dm}/64\simeq 3.53\,{\rm M}_\odot$ and $M_{\rm gas}=(\Omega_b/\Omega_m)M_{\rm dm}/64\simeq 0.72\,{\rm M}_\odot$, respectively. We use a comoving gravitational softening length of $17\,{\rm pc}$ for the refined DM component.

\subsection{The Moving Mesh Code {\small AREPO}}

We follow the collapse of the gas in the refined minihalos with the cosmological moving mesh code {\small AREPO} \citep{springel10a}. {\small AREPO} is a second-order accurate finite volume method that solves the Euler equations based on a piece-wise linear reconstruction and the calculation of hydrodynamical fluxes at every cell face with an exact Riemann solver. The principle difference of {\small AREPO} compared to Eulerian mesh codes is that its computational mesh is constructed as the Voronoi tessellation of a set of mesh-generating points. These points can be moved with the flow velocity itself, making the mesh automatically adaptive in a Lagrangian fashion. This technique greatly reduces the numerical diffusivity of mesh-based hydrodynamics, especially when large bulk flows are present. In fact, the results of the code become fully Galilean-invariant, whereas the truncation error of ordinary Eulerian mesh codes depends on the bulk velocity of the system. In addition, the unstructured Voronoi mesh of {\small AREPO} avoids the introduction of preferred directions, which are present in Cartesian meshes.

The novel {\small AREPO} scheme hence combines the accuracy of mesh-based hydrodynamics with the natural adaptivity and translational invariance usually only provided by the smoothed particle hydrodynamics (SPH) technique. In terms of hydrodynamical accuracy, the grid-based approach of {\small AREPO} alleviates a number of shortcomings encountered with SPH \citep{monaghan05}. Among these are the inherent noise of the kernel estimates, the artificial viscosity, and the slow convergence rate of SPH in three dimensions \citep{springel10b}. Further important improvements of {\small AREPO} lie in the more accurate treatment of shocks and turbulence, and of fluid instabilities \citep{agertz07}. Compared to adaptive mesh refinement (AMR) codes, a significant advantage of {\small AREPO} lies in its ability to continuously adjust its resolution when density fluctuations grow under self-gravity. This key feature of Lagrangian codes is ideal for gravitational collapse problems. If the bulk velocities are large, {\small AREPO} can use larger timesteps than AMR codes and exhibits lower numerical diffusion errors at comparable spatial resolution.

We note that the computational speed of {\small AREPO} in simulations of cosmic structure formation is roughly on par with {\small GADGET-3} for an equal number of hydrodynamic resolution elements, despite its comparatively complex mesh-construction calculations. This is in part due to the large cost of accurate calculations of self-gravity, which dominate over the hydrodynamics both in {\small AREPO} and {\small GADGET-3}. Both codes compute the gravitational field with a fully adaptive combination of a particle-mesh (PM) solver for large-scale forces, and a hierarchical multipole expansion for short-range forces. In any case, because of the higher accuracy of {\small AREPO} compared to SPH for an equal number of resolution elements, this means that {\small AREPO} is the more efficient method. The faster convergence rate of {\small AREPO} also implies that its accuracy advantage increases further if more resolution elements are employed.

The mesh-generating points of {\small AREPO} can still be interpreted as Lagrangian fluid particles, if desired, which greatly simplifies the (re)use of well-tested techniques from SPH codes, such as implementations of gas chemistry or sink particles, an approach we employ extensively in this study. Also, even though {\small AREPO}'s fluid cells follow the motion of the gas and hence maintain a roughly constant gas mass, it is sometimes necessary to refine the mass resolution per cell during a calculation. This can be done accurately in the mesh-based approach of {\small AREPO}, as we discuss next.

\subsection{Mesh Refinement}

It is well established that an essential prerequisite for reliable hydrodynamic collapse simulations lies in a numerical resolution of the Jeans length. In SPH simulations, the Jeans mass associated with each particle must exceed the mass within the smoothing length by a factor of about two \citep{bb97}, while grid-based codes rely on the so-called `Truelove criterion' for satisfying this requirement \citep{truelove98}, which stipulates that in an isothermal collapse calculation at least four cells per Jeans length should be used. Collapse simulations may therefore require additional run-time refinement to ensure that the Jeans length is resolved and artificial fragmentation is avoided.

The addition of cells in grid codes is comparatively straightforward, while special care must be taken in SPH simulations to avoid substantial noise when new particles are inserted \citep{kw02}. In {\small AREPO}, a very simple and robust approach exists: if a cell fulfils a predefined refinement criterion, it is split by inserting a further mesh-generating point at the position of the generator of the original cell, displaced by a random offset that is very small compared to the size of the cell.  The mass, momentum and energy of the original cell are then distributed conservatively among the two new cells, weighted by their respective volumes, keeping the density, velocity and pressure of the original cell unchanged. Over the course of a few timesteps, the code then separates the mesh-generating points of the two cells and moves them closer to their geometric centers of mass, as a result of the mesh-steering corrections employed by {\small AREPO} to maintain a regular structure of the mesh. We note that the numerical fluxes required for this mesh readjustment arise consistently from the Riemann solutions calculated by the code, as these take the mesh motion fully into account.

We here briefly investigate how the resolution of the Jeans length affects the amount of artificial fragmentation found in {\small AREPO}, using a series of isothermal collapse simulations in the formulation of \citet{bb93}, which is based on a test initially introduced by \citet{bb79}. This setup consists of a $1\,{\rm M}_\odot$ cloud with radius $r=5\times 10^{16}\,{\rm cm}$ and sound speed $c_{\rm s}=1.66\times 10^4\,{\rm cm}\,{\rm s}^{-1}$ in solid-body rotation with an angular velocity of $\omega =7.2\times 10^{-13}\,{\rm rad}\,{\rm s}^{-1}$ and an $m=2$ density perturbation of the form
\begin{equation}
\rho(\phi)=\rho_0[1+0.1\,{\rm cos}(2\phi)]\mbox{\ ,}
\end{equation}
where $\rho_0=3.82\times 10^{-18}\,{\rm g}\,{\rm cm}^{-3}$ is the underlying constant density and $\phi$ is the azimuthal angle around the rotation axis. We refine the gas whenever the local Jeans number, here defined as the radius of a cell divided by the local Jeans length, increases above a predefined value. We estimate the cell radius as $h=(3V/4\pi)^{1/3}$, where $V$ is the volume of the cell. Since the mesh-steering motions ensure that the cells do not become too distorted, this provides a good estimate of the size of a cell. In Figure~1, we show the state of the gas in one of the two main clumps for a minimum of one, two, four, and eight cells per Jeans length. For isothermal gas, artificial fragmentation sets in above a Jeans number of approximately $1/4$, which is similar to the result found in AMR codes \citep{truelove98}. However, this should be considered a lower limit on the required resolution.

In the present study, we activate the refinement criterion in our cosmological simulations above a density of $n_{\rm H}=1\,{\rm cm}^{-3}$ in the central $200\,{\rm pc}$ of the halo, which is larger than the maximum virial radius of the minihalos investigated here (see Table~1). We ensure that the local Jeans length is resolved by at least $128$ cells until the number density of hydrogen atoms exceeds $n_{\rm H}=10^9\,{\rm cm}^{-3}$, at which point we extract the central $1\,{\rm pc}$ of the simulations and deactivate further refinement. We reinitialize the simulations in this smaller box with reflective boundary conditions \citep[see][]{springel10a}, which ensures that the gas remains confined by external pressure. We also discard all DM particles, since the gravitational potential at these densities is dominated by the gas: the radius at which the gas mass exceeds the DM mass by an order of magnitude is $0.76$, $0.67$, $0.58$, $0.71$ and $1.1\,{\rm pc}$, respectively, which is comparable to the spatial extent of the resimulations. Due to the deactivation of the refinement, the particle mass within the central $\simeq 1000\,{\rm AU}$ remains roughly constant at $10^{-4}\,{\rm M}_\odot$. We note that shocks from inflows into the innermost region are still well resolved under these conditions, thanks to the ability of {\small AREPO} to capture shocks on $\sim 1-2$ cells. Assuming a temperature of $1000\,{\rm K}$, the Truelove criterion is thus expected to be violated at a density of $\simeq 10^{17}\,{\rm cm}^{-3}$, which is approximately the density at which the sink particles are inserted.

\subsection{Sink Particles}

The evolution of the gas beyond the runaway collapse of the first protostellar core is captured with a sink-particle algorithm that exploits the hybrid nature of {\small AREPO}. Since the mesh-generating points are Lagrangian tracers of the fluid, sink particles may be treated very similarly to implementations commonly used in SPH simulations \citep{bbp95,bcl02,jappsen05}. However, instead of using a density threshold, we identify candidate sink particles whenever the Jeans number of a cell falls below a predefined critical Jeans number $J_{\rm crit}=1/8$. For a sink particle to form, the gas within $r_{\rm sink}=h/J_{\rm crit}$ is further required to be bound and have a negative velocity divergence. All mesh-generating points within $r_{\rm sink}$ are then removed and replaced by a collisionless sink particle with a gravitational softening length of $r_{\rm acc}/3$, where $r_{\rm acc}$ is a predefined accretion radius that is set independently from the initial sink particle radius. In practice, only $\simeq 200$ mesh-generating points are removed during this step, since the spatial resolution around the sink particle decreases at larger radii. The sink particle is placed at the center of mass of the removed cells with a velocity determined by linear momentum conservation, while the angular momentum and internal energy of the gas are discarded. The additional factor $1/3$ in the gravitational softening is used to avoid artificial fragmentation, which might occur if gravitational forces on the gas are reduced on the scale of the accretion radius.

Accretion onto existing sink particles occurs if the mesh-generating point associated with a candidate cell falls within the accretion radius $r_{\rm acc}$ of the sink particle it is most bound to. This method exploits the Lagrangian nature of the mesh-generating points, and yields the (incremental) rate at which mass flows onto sink particles. We have also tested an implementation with a more stringent criterion, where in addition to being bound to the sink particle, the semimajor axis of the respective two-body system must fall below the accretion radius. However, we have found that these additional criteria made no appreciable difference in practice.

Once a cell has been accreted, its mass and momentum is added to the sink particle and the corresponding mesh-generating point is removed. After this step has been performed for all candidate cells, the mesh is reconstructed such that the volume associated with the removed cells is distributed among the remaining cells around the sink particle. As conserved quantities are weighted with the new volumes, this tends to make their densities and pressures artificially small. In the last section, we present a resolution study with varying accretion radii to show that this caveat artificially reduces the amount of fragmentation, and increases the typical fragment mass. As our fiducial accretion radius, we choose a value of $r_{\rm acc}=100\,{\rm R}_\odot$, which is close to the maximum physical size of accreting Pop~III stars \citep{ho09}.

Mergers between sink particles occur whenever the total energy of the respective two-body system is negative and the semimajor axis falls below the accretion radius. This is motivated by our above choice, which yields an approximate upper limit on the physical size of the protostars. In cases where more than one sink particle lies within the accretion radius of another sink particle, the pair with the highest binding energy merges. For comparison, we have also implemented an alternative merging criterion, where sink particles merge whenever their separation drops below $r_{\rm acc}$, and no further checks on their energies are enforced. The purpose of this formulation is to maximize the efficiency of merging between protostars, since we do not capture the gasdynamical friction between real protostars.

\subsection{Chemistry and Cooling}

The chemical and thermal evolution of the gas is modeled as described in \citet{clark11a}. This network consists of $45$ reactions amongst twelve chemical species: H, H$^+$, H$^-$, H$_2^+$, H$_2$, He, He$^+$, He$^{++}$, D, D$^+$, HD, and free electrons. We account for all relevant cooling processes, including electronic excitation of H, He and He$^+$, cooling from the recombination of H$^+$ and He$^+$, Compton cooling, and bremsstrahlung. In practice only a few processes play an important role at the densities and temperatures investigated here -- H$_2$ rotational and vibrational line cooling, H$_2$ collision-induced emission (CIE) cooling, H$_2$ collisional dissociation cooling, and heating due to three-body H$_2$ formation \citep[see also][]{turk11}.

Collisional dissociation of H$_2$ is modeled with the rate given by \citet{msm96}, from which the three-body formation rate is derived by applying the principle of detailed balance \citep{fh07,glover08}. These rates are intermediate in terms of the rates discussed in \citet{turk11}, and their uncertainty is reflected in the substantial variation of the thermal and morphological evolution of primordial gas clouds. In the present study, we do not investigate how the collapse of the gas is affected by our choice of the three-body rate. However, we note that the rate used by \citet{clark11b}, taken from \citet{abn02}, is significantly smaller than the rate adopted here, and the fact that we see qualitatively similar behavior in both simulations gives us confidence that our main results do not depend to any great extent on the choice of three-body rate, although the quantitative details will likely have some dependence on the choice of rate.

The cooling of the gas by H$_2$ lines in the optically thin regime is modeled with the low-density cooling rates for collisions between H$_2$ molecules and H and He atoms, H$_2$ molecules, protons and electrons \citep{ga08}, accounting for the transition to local thermodynamic equilibrium level populations at gas densities $n_{\rm H}\gg 10^{4}\,{\rm cm}^{-3}$. At densities above $n_{\rm H}\sim 10^{9}\,{\rm cm}^{-3}$, the strongest of the H$_2$ ro-vibrational lines become optically thick, reducing the effectiveness of H$_2$ line cooling. To account for this effect, we use an approach based on the Sobolev approximation \citep{yoshida06b}. We write the H$_2$ cooling rate as
\begin{equation}
\Lambda_{{\rm H}_2}=\sum_{\rm u,l}\Delta E_{\rm ul}A_{\rm ul}\beta_{\rm esc, ul}n_{\rm u}\mbox{\ ,}
\end{equation}
where $n_{\rm u}$ is the number density of hydrogen molecules in upper energy level $u$, $\Delta E_{\rm ul}$ is the energy difference between this upper level and a lower level $l$, $A_{\rm ul}$ is the spontaneous radiative transition rate for transitions between $u$ and $l$, and $\beta_{\rm esc, ul}$ is the escape probability associated with this transition, i.e. the probability that the emitted photon can escape from the region of interest. We fix $n_{\rm u}$ by assuming that the H$_2$ level populations are in local thermodynamic equilibrium, and write the escape probabilities for the various transitions as
\begin{equation}
\beta_{\rm esc, ul}=\frac{1-\exp(-\tau_{\rm ul})}{\tau_{\rm ul}}\mbox{\ ,}
\end{equation}
where we use the approximation that
\begin{equation}
\tau_{\rm ul}\simeq\alpha_{\rm ul}L_{\rm s}\mbox{\ ,}
\end{equation}
where $\alpha_{\rm ul}$ is the line absorption coefficient and $L_{\rm s}$ is the Sobolev length \citep{yoshida06b}. In the classical, one-dimensional spherically symmetric case, the Sobolev length is given by
\begin{equation}
L_{\rm s}=\frac{v_{\rm th}}{|{\rm d}v_{\rm r}/{\rm d}r|}\mbox{\ ,}
\end{equation}
where $v_{\rm th}$ is the thermal velocity, and ${\rm d}v_{\rm r}/{\rm d}r$ is the radial velocity gradient. In our three-dimensional simulations, we generalize this as \citep{nk93}
\begin{equation}
L_{\rm s}=\frac{v_{\rm th}}{|\nabla\cdot {\mathbf v}|}\mbox{\ .}
\end{equation}
To prevent the H$_2$ cooling rate from being reduced by an unphysically large amount in regions with small velocity gradients, we limit $L_{\rm s}$ to be less than or equal to the local Jeans length, $L_{\rm J}$. We justify this choice by noting that there are strong density gradients in the gas on scales $L\gg L_{\rm J}$, and so we expect the bulk of the contribution to the H$_2$ line absorption to come from material within only a few Jeans lengths.

At very high densities ($n_{\rm H}>10^{14}\,{\rm cm}^{-3}$), CIE cooling from H$_2$ becomes more effective than H$_2$ line cooling \citep{on98,ra04}. We account for the reduction of the CIE cooling rate due to the effects of continuum opacity using the following prescription \citep{ripamonti02,ra04}:
\begin{equation}
\Lambda_{\rm CIE, thick}=\Lambda_{\rm CIE, thin}\,{\rm min}\left(\frac{1-e^{-\tau_{\rm CIE}}}{\tau_{\rm CIE}},1\right)\mbox{\ ,}
\end{equation}
where
\begin{equation}
\tau_{\rm CIE}=\left(\frac{n_{\rm H_2}}{7\times 10^{15}\,{\rm cm}^{-3}}\right)^{2.8}\mbox{\ .}
\end{equation}
Finally, we account for the fact that each time an H$_2$ molecule is collisionally dissociated, $4.48\,{\rm eV}$ of thermal energy is lost by the gas, while every time that a new H$_2$ molecule is formed by the three-body process, the gas gains $4.48\,{\rm eV}$.

\section{Results}

We here discuss the collapse of the gas in the minihalos up to the formation of the first protostar, and the subsequent fragmentation and accretion that leads to the build-up of the protostellar cluster. We then proceed to investigate the influence of radiation, present a resolution study, and discuss the caveats of the sink-particle algorithm.

\subsection{Collapse of Gas in Minihalos}

In Figures~2 and 3, we show the collapse of the gas in all five minihalos from cosmological to protostellar scales. As gas falls into the DM halo, shocks thermalize the gravitationally generated infall motions and heat the gas to the halo virial temperature. At the same time, irregular motions of the dark matter seed subsonic turbulence in the gas, which cascades down to ever smaller scales and later influences the fragmentation of the protostellar cloud. Once enough hydrogen molecules have formed, the gas at the center of the halo begins to cool, becomes gravitationally unstable, and decouples from the dark matter on a scale of a parsec \citep{abn02,bcl02}. At significantly higher densities, three-body reactions render the gas fully molecular in a region of size a few hundred AU \citep{yoshida06b}. Somewhat later, the gas becomes optically thick and the cooling radiation can no longer escape, inducing a final heating phase that ends with the formation of a protostar \citep{yoh08}. At this point a sink particle with an accretion radius of $100$ solar radii is inserted, which is the maximum photospheric size of the accreting protostar \citep{ho09}.

Although the evolution of the gas is quite similar in all cases, some scatter in the thermal history exists \citep[e.g.,][]{on07}. In simulations MH-2 and MH-3, HD cooling becomes important and leads to a prolonged cooling phase at densities $n_{\rm H}\ga 10^4\,{\rm cm}^{-3}$ \citep{ripamonti07,mb08}. This is evident from Figures~4 and 5, where we show the occupation of the gas in density and temperature space. There is no correlation of the prolonged cooling with the bulk properties of the underlying DM halo (see Table~1), indicating that details of the virialization process must be responsible for this effect. Note that the temperature floor required for HD cooling is around $150\,{\rm K}$, which is only marginally attained in most minihalos. The activation of HD cooling is therefore very sensitive to the evolution of the gas during the initial free-fall phase. A second interesting effect is the shock-heating of some parcels of gas, which we attribute to the more violent heating in simulations MH-2 and MH-3 once three-body reactions set in. A peak similar to the one found at $n_{\rm H}\sim 10^9\,{\rm cm}^{-3}$ in simulation MH-2 was also found in \citet{tna10}, although it was much more pronounced.

In Figure~6, we show the internal structure of the minihalos just before the formation of the first protostar. The density, enclosed mass, and angular momentum profiles are relatively similar and agree well with previous investigations \citep{abn02,yoshida06b}. The second density peak in simulation MH-2 indicates the nearly simultaneous collapse of a second gas cloud, which was also found in \citet{tao09}. The temperature profiles show significant scatter due to the previously mentioned differences in the virialization of the individual minihalos. We find no convincing correlation of the thermal history of the gas with the velocity dispersion. The latter is always close to Mach numbers $M\simeq 1$, indicating transsonic turbulence.

\begin{figure*}
\begin{center}
\resizebox{13.5cm}{10.5cm}
{\unitlength1cm
\begin{picture}(13.5,10.5)
\put(0,6){\includegraphics[width=4.5cm,height=4.5cm]{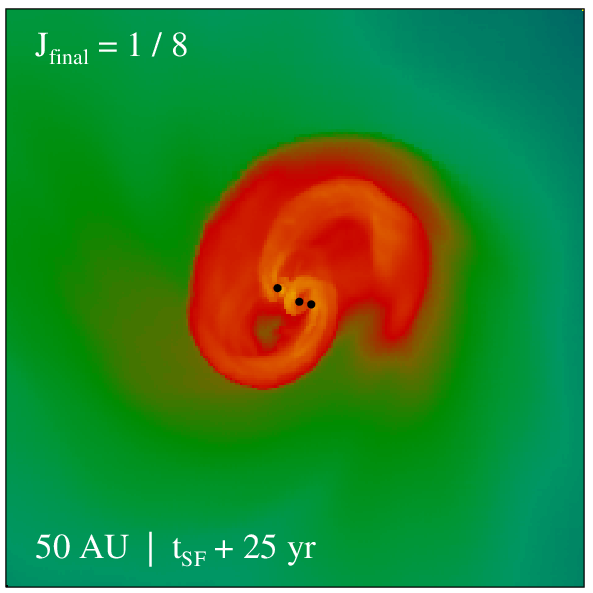}}
\put(4.5,6){\includegraphics[width=4.5cm,height=4.5cm]{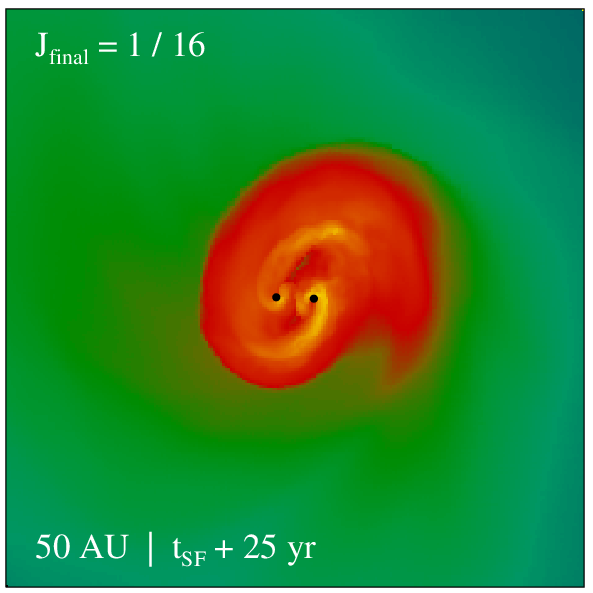}}
\put(9,6){\includegraphics[width=4.5cm,height=4.5cm]{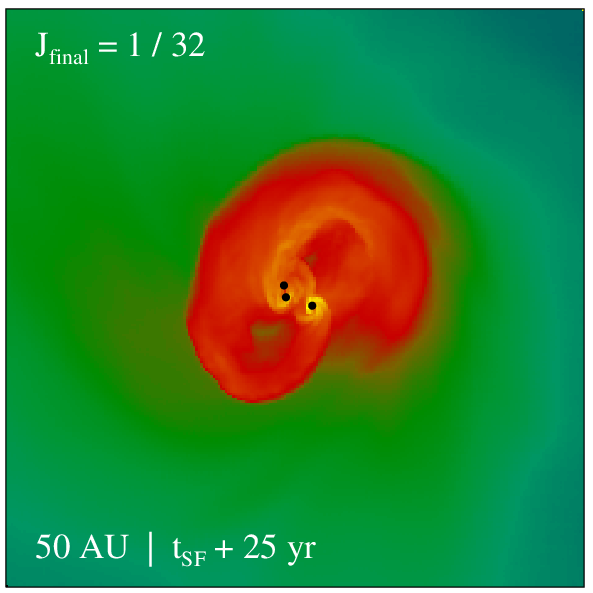}}
\put(0,1.5){\includegraphics[width=4.5cm,height=4.5cm]{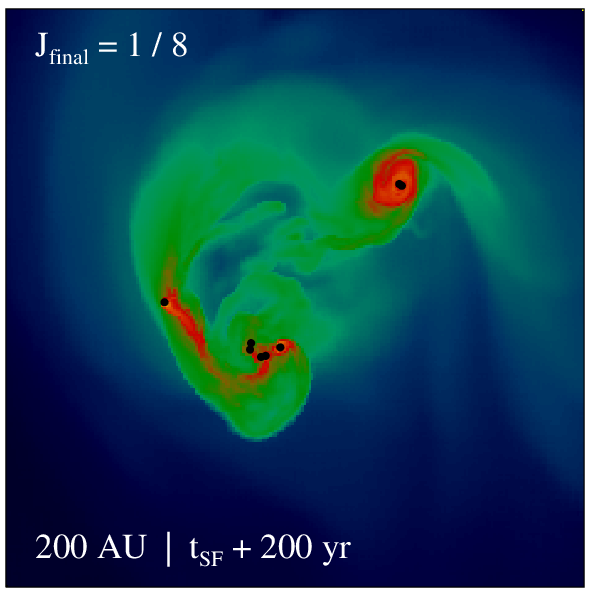}}
\put(4.5,1.5){\includegraphics[width=4.5cm,height=4.5cm]{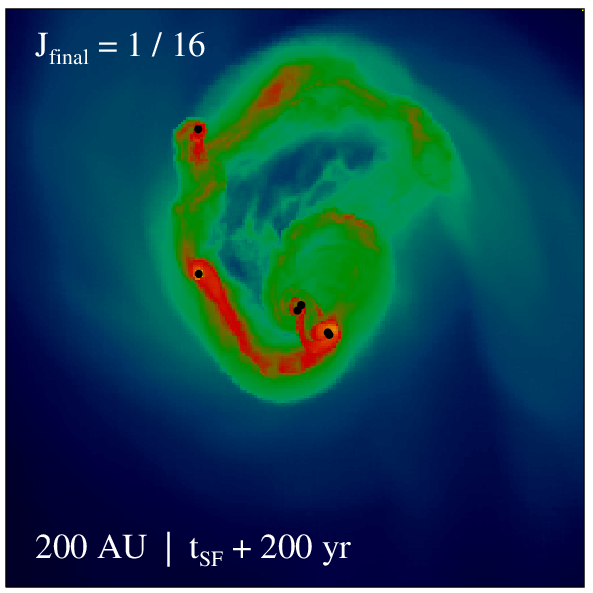}}
\put(9,1.5){\includegraphics[width=4.5cm,height=4.5cm]{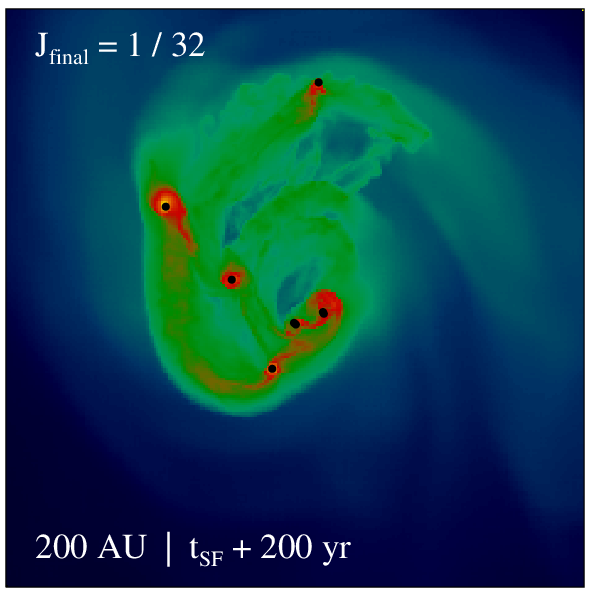}}
\put(0,0){\includegraphics[width=13.5cm,height=1.5cm]{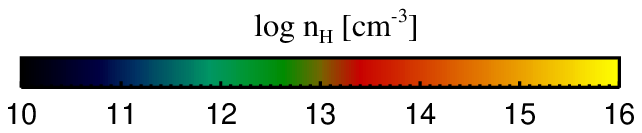}}
\end{picture}}
\caption{Resolution study of simulation MH-1 using standard sink particles. The top panels show the initial fragmentation of the disk in a box with $50\,{\rm AU}$ per side after $25\,{\rm yr}$ and for $8$, $16$ and $32$ cells per Jeans length. The individual simulations are very similar during the initial formation and fragmentation of the disk, with the exception that in the middle panel two of the protostars have just merged. At later times the simulations begin to diverge quite strongly due to chaotic N-body interactions between the protostars. As shown in the bottom panels, the morphologies have become quite distinct after $200\,{\rm yr}$, even though the total numbers of protostars are nearly equal.}
\end{center}
\end{figure*}

\begin{figure*}
\begin{center}
\includegraphics[width=14cm]{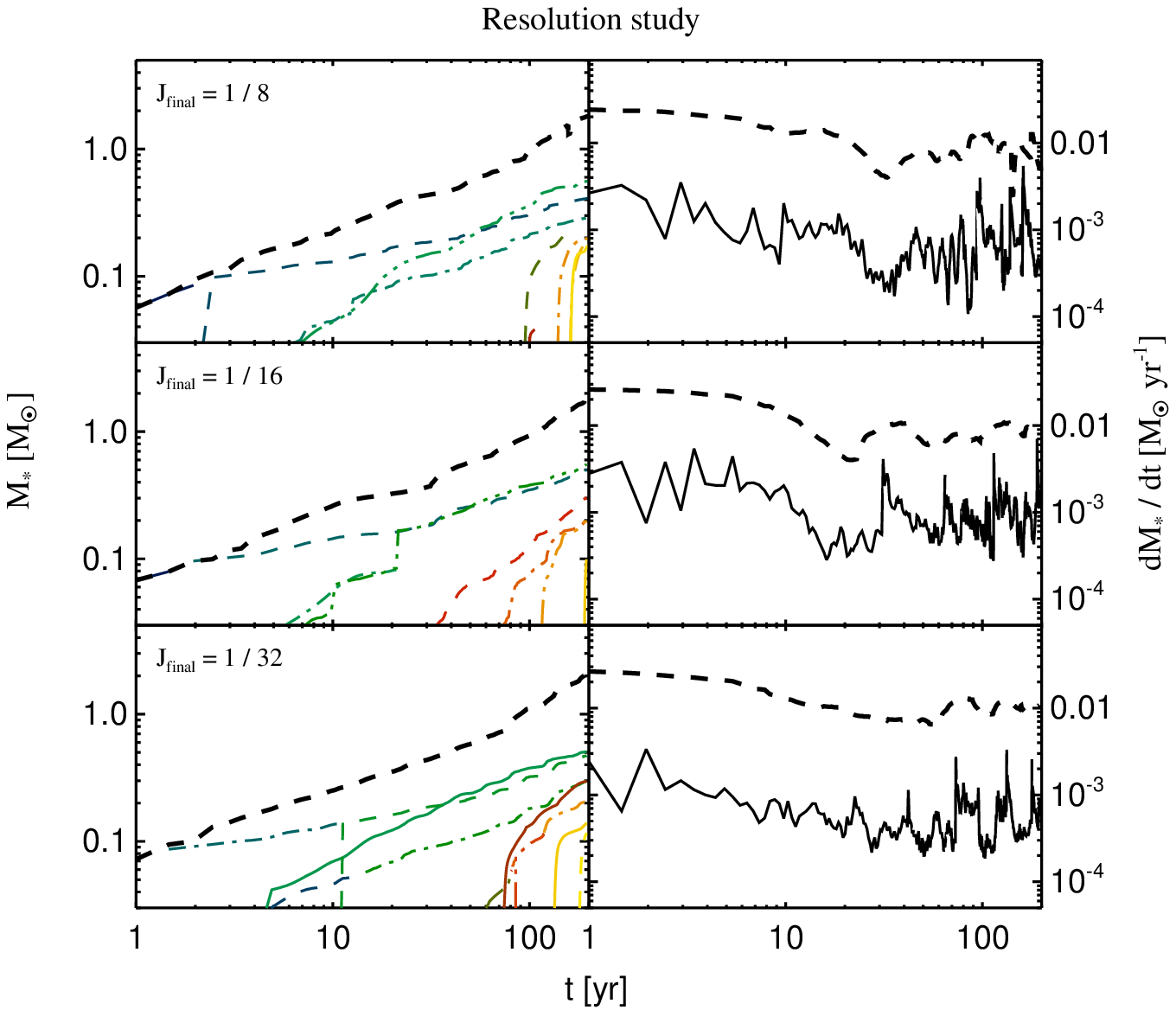}
\caption{Same as Figure~9, but for the resolution study of simulation MH-1. The three runs show some differences in the mass accretion histories of individual protostars due to the chaotic nature of N-body interactions. However, their numbers and total mass, as well as their cumulative and average accretion rates agree well with each other. The employed resolution therefore does not qualitatively affect the fragmentation of the gas.}
\end{center}
\end{figure*}

\subsection{Fragmentation and Protostellar Mass Function}

Shortly after the formation of the first protostar, the residual angular momentum of the surrounding gas cloud leads to the formation of a circumstellar accretion disk \citep{clark11b}. Since the infall rate of new material onto the disk is extremely high, angular momentum transport by viscous and gravitational torques is not sufficient to prevent the disk from fragmenting. As shown in Figure~7, the disk hence passes the gravitational instability limit, at which point it forms a number of secondary protostars. After this initial burst, the gas becomes gravitationally unstable multiple times and forms a small cluster of $\sim 10$ protostars after only $1000\,{\rm yr}$ (see Figure~8).

The complex gravitational interactions between individual protostars and the surrounding gas clouds are illustrated in Figures~9 and 10, where we show the mass accretion histories of the protostars for two different sink-particle schemes. In our standard approach, sink particles merge if they become gravitationally bound and pass within $100$ solar radii of each other, while in our `adhesive' approach, only the latter criterion must be fulfilled. The second model is a crude representation of the dissipative effects of shocks in collisions between physically extended protostars. Both techniques have been used in (radiation) hydrodynamic simulations of star cluster formation in the present-day universe \citep{kb00,bbb03,krumholz09,federrath10,peters10}. In the absence of a detailed understanding of the true protostellar radii and the influence of tidal forces during close encounters, we hope that they bracket physical reality.

Interestingly, the two methods predict nearly identical total protostellar masses, but yield different detailed mass accretion histories. For standard sink particles, protostars generally survive close encounters, whereas for adhesive sink particles these interactions often result in a merger. For this reason the number of protostars after $1000\,{\rm yr}$ is higher in our standard formulation of sink particles than in our adhesive formulation. This is evident from Figure~11, where we show the final protostellar mass functions of all five minihalos. Notably, a small group of protostars with a range of masses is formed in both cases. The cumulative protostellar mass functions obtained from the five independent realizations are relatively flat, extending from $\sim 0.1\,{\rm M}_\odot$ to nearly $10\,{\rm M}_\odot$ (bottom right panel of Figure~11). This corresponds to a top-heavy mass function, in the sense that most of the protostellar mass is locked up in massive protostars.

An interesting trend obtained in the case of standard sink particles is that some low-mass protostars are ejected from the central gas cloud by dynamical interactions with more massive protostars, while the latter tend to remain at the center of the minihalo and continue to accrete aggressively. This is also visible in Figure~12, where we show the radial velocity and ratio of the radial velocity to the escape velocity of all protostars. A number of protostars move away from the center of mass with speeds exceeding the escape velocity, such that they stop accreting and might enter the main sequence as low-mass Pop~III stars. This still occurs in our formulation of adhesive sink particles, but less frequently. Furthermore, in this case no protostars reside within the central $\simeq 100\,{\rm AU}$, since momentum conservation acts to move merged protostars to larger annuli.

\subsection{Radiation Feedback}

Studies of present-day star formation have shown that the radiation emitted by young protostars heats the surrounding envelope and reduces the degree of fragmentation near the central protostar \citep{krumholz09,peters10}. However, the outcome is less clear in the present study, where gas temperatures are a factor of $10-100$ times higher and different cooling mechanisms are in play. To investigate the importance of radiation in detail, we compare simulations with and without the heating from accreting Pop~III stars in one of the five minihalos.

During the early stages of protostellar evolution, Pop~III stars grow rapidly in size from $\simeq 10$ to $\simeq 100\,{\rm R}_\odot$ \citep{ho09}. The temperature of the photosphere remains roughly constant at $6000\,{\rm K}$, and so most of the energy is radiated in the optical and infrared. Once the protostar has accreted of order $10\,{\rm M}_\odot$, the star undergoes Kelvin-Helmholtz contraction and heats up substantially, leading to the emission of ionizing radiation. Since we are well within the initial swelling phase, we can utilize a simple optically thin treatment of the radiation to determine the maximum extent to which the gas will be affected.

\begin{figure*}
\begin{center}
\includegraphics[width=14cm]{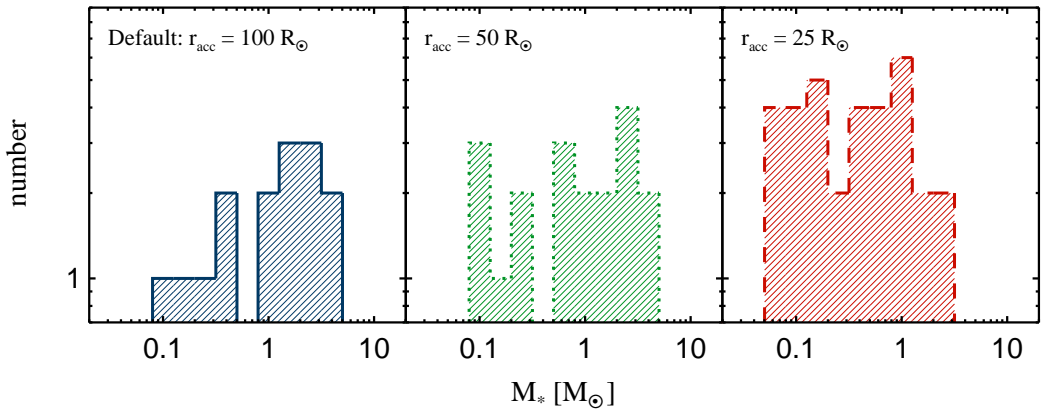}
\caption{The final mass function obtained in simulation MH-1 using accretion radii decreased by factors of two and four from the fiducial value $r_{\rm acc}=100\,{\rm R}_\odot$. The collapse of the gas is thus followed to higher densities and smaller separations from the protostar. Any systematic inaccuracies stemming from the artificially reduced density and pressure around sink particles should therefore induce a trend. As is evident from the figure, such a trend exists: the amount of fragmentation increases, while the typical mass of a fragment decreases. This shows that our results should be considered a lower limit on the degree of fragmentation in minihalos.}
\end{center}
\end{figure*}

For this purpose we treat each sink particle formed in our simulation as a separate protostar, and account for the energy released by accretion onto the surfaces of these protostars. To model the effects of this accretion luminosity, we first compute the bolometric accretion luminosity for each protostar
\begin{equation}
L_{\rm acc}=\frac{G{\dot M}_*M_*}{R_*}\mbox{\ ,}
\end{equation}
where ${\dot M}_*$ is the accretion rate onto the protostar, $M_*$ is the protostellar mass and $R_*$ is the protostellar radius. We relate the protostellar radius to the current protostellar mass and accretion rate using a relationship derived for adiabatically accreting, metal-free protostars embedded in a spherically symmetric inflow \citep{sps86}:
\begin{equation}
R_*=26\,{\rm R}_\odot\left(\frac{M_*}{{\rm M}_\odot}\right)^{0.27}\left(\frac{{\dot M}_*}{10^{-3}\,{\rm M}_\odot\,{\rm yr}^{-1}}\right)^{0.41}\mbox{\ .}
\end{equation}
More recent calculations for the case of rotating infall find an even larger value for $R_*$ during the adiabatic accretion phase \citep{tm04}, and so using the above expression gives us a conservative upper limit on the bolometric accretion luminosity during this phase.

Combining the above equations, we find that
\begin{equation}
 L_{\rm acc}\simeq 1200\,{\rm L}_\odot\left(\frac{M_*}{{\rm M}_\odot}\right)^{0.73}\left(\frac{{\dot M}_*}{10^{-3}\,{\rm M}_\odot\,{\rm yr}^{-1}}\right)^{0.59}\mbox{\ .}
\end{equation}
With this expression, we determine the heating rate of the gas surrounding the protostar from
\begin{equation}
\Gamma_*=\rho\kappa_{\rm p}\frac{L_{\rm acc}}{4\pi r^2}\mbox{\ ,}
\end{equation}
where $\rho$ is the mass density, $r$ is the distance to the protostar, and $\kappa_{\rm p}$ is the Planck mean opacity of the gas. We calculate this mean opacity by interpolation, using tabulated values that account for both line and continuum absorption \citep{md05}, and that include the influence of the electrons provided by ionized lithium. This expression assumes that the gas is optically thin to the radiation from the accreting protostar. Making this assumption allows us to avoid the extremely high computational cost that would be associated with an accurate treatment of the transfer of the protostellar radiation, and also gives us a conservative upper limit on the effectiveness of protostellar feedback. Finally, we assume that each protostar accretes at a rate of $0.1\,{\rm M}_\odot\,{\rm yr}^{-1}$, which is well above the total accretion rate of the whole ensemble of protostars, maximizing the effect of our radiative feedback model.

In Figure~13, we compare runs with and without radiation feedback for simulation MH-1. We find that the additional heating results in a smoother and more extended disk, in which the fragments form at slightly larger distances from the center \citep[see also][]{clark11b}. However, even the unrealistically strong radiation field assumed here does not prevent fragmentation within the disk. The primary reason for this is the strong cooling of the gas by molecular hydrogen lines at temperatures around $1000\,{\rm K}$ and above, which largely offsets the heating caused by the accretion luminosity \citep[e.g.,][]{omukai10}. The neglect of radiation in our simulations should therefore not qualitatively affect our conclusions. We note that a more detailed investigation of radiation feedback for the minihalos used here is presented in a companion paper \citep{smith11}. Finally, it is important to point out that the heating may be much more severe once the first protostar has accreted of order $10\,{\rm M}_\odot$ and begins to emit ionizing radiation \citep{tm04,mt08}. However, this will not occur until after the period of time simulated here.

\subsection{Resolution Study}
We here study how our results depend on the resolution employed. For this purpose we rerun simulation MH-1 with a modified refinement criterion. We linearly interpolate the number of cells per Jeans length between two threshold densities, using $J_{\rm init}=1/128$ at a density of $n_{\rm H}=10^9\,{\rm cm}^{-3}$ and $J_{\rm final}$ at $n_{\rm H}=10^{17}\,{\rm cm}^{-3}$. This allows a better control of the number of resolution elements at a given density, and thus on the scale where the sink particles are created. We run three simulations with inverse Jeans numbers of $J_{\rm final}=1/8, 1/16$ and $1/32$, corresponding to factors of $8$ increase in the number of particles at $n_{\rm H}=10^{17}\,{\rm cm}^{-3}$. In analogy to our main simulations, sink particles are created once $J_{\rm crit}=J_{\rm final}$, with the exception that we here use an initial sink particle radius equal to the accretion radius of $100\,{\rm R}_\odot$.

The mass resolution of these simulations is significantly higher than in our original simulations, which greatly increases their computational cost. The mass in the cells at the highest densities has dropped by nearly two orders of magnitude to $\simeq 10^{-6}\,{\rm M}_\odot$, so that we have only investigated the first $200\,{\rm yr}$ instead of the original $1000\,{\rm yr}$. However, this period of time is sufficient to show that the overall fragmentation process is not affected by the resolution employed. In Figure~14, we show the central $50\,{\rm AU}$ of all three simulations after $25\,{\rm yr}$. The morphologies of the disk and the positions of the protostars agree well with each other, with the exception that in the case with $J_{\rm final}=1/16$ one of the protostars has just merged. Somewhat later, N-body interactions lead to an increasingly chaotic evolution of the system, so that small differences in the initial conditions result in a relatively large dispersion in the evolutionary paths of individual protostars. This is evident from the bottom panels, where we show the protostellar cluster in a box with $200\,{\rm AU}$ on a side and after $200\,{\rm yr}$. The appearance of the cluster has become quite distinct in each case. However, Figure~15 shows that the numbers of protostars and their total mass are much more insensitive to the initial conditions. The employed resolution can thus change the details of the fragmentation due to a `butterfly effect', but not qualitatively affect our results.

\subsection{Variation of Accretion Radius}

The removal of mesh-generating points below the accretion radius artificially reduces the density and pressure around sink particles. We investigate how this affects our results by systematically decreasing the accretion radius from $100\,{\rm R}_\odot$ to $50\,{\rm R}_\odot$ and $25\,{\rm R}_\odot$. The gas therefore collapses to progressively higher densities before being accreted onto the sink particle, such that any residual effects should display a trend in the total number of fragments formed, as well as on the masses of the fragments. As is evident from Figure~16, such a trend exists: A reduction of the accretion radius results in an increase in the number of fragments and a decrease of the typical fragment mass. Existing sink particles apparently accrete less aggressively, such that more mass remains in the disk, which in turn becomes even more susceptible to fragmentation. This trend indicates that our results should be considered a lower limit on the degree of fragmentation in minihalos.

\subsection{Protostellar Merging}

Another limitation of the sink-particle algorithm employed here is that the gasdynamical friction between physically extended protostars is not modeled. Mergers between protostars may therefore be more common than for the purely gravitationally interacting sink particles employed here. To understand the potential importance of gasdynamical friction, we use Equation~10 to compare the minimum separation of each protostar to any other protostar with the sum of the protostellar radii during their closest encounter. Figure~17 shows that the majority of protostars experience at least one close encounter with another protostar and, depending on the eccentricity and infall velocity of the respective orbit, will merge. Only a few low-mass protostars remain physically separated for the entire duration of the simulation, while acquiring substantial radial velocities (see Figure~12). As mentioned already in Section~3.2, these protostars stop accreting and may therefore enter the main sequence as low-mass Pop~III stars. However, this possibility depends sensitively on our choice of the accretion radius, which might be too small considering that real protostars are likely surrounded by an unresolved accretion disk. Assuming a somewhat larger accretion radius would further reduce the number of low-mass protostars that could survive. We note that an accurate account of merging between protostars would require simulations that self-consistently capture the interaction of the protostars with the surrounding gas cloud. These are not currently feasible.

\begin{figure}
\begin{center}
\includegraphics[width=8cm]{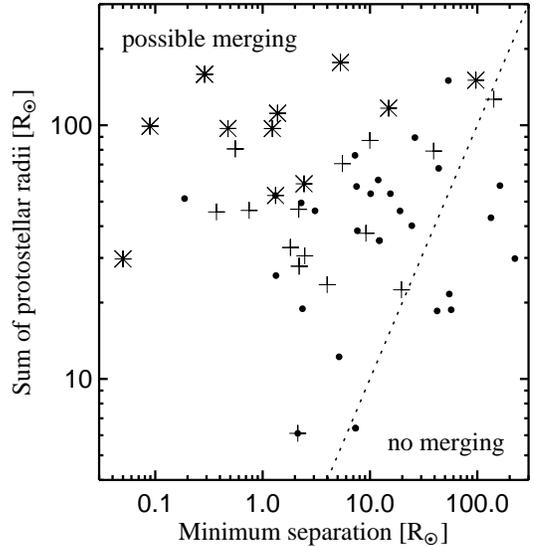}
\caption{A comparison of the minimum separation of each protostar to any other protostar with the sum of the protostellar radii during their closest encounter. Black dots, crosses and stars denote protostars with masses below $1\,{\rm M}_\odot$, between $1\,{\rm M}_\odot$ and $3\,{\rm M}_\odot$, and above $3\,{\rm M}_\odot$. Nearly all protostars experience at least one encounter with another protostar where the separation drops below their total physical size. Depending on the eccentricity and infall velocity of the respective orbits, they will merge. As delineated by the dotted line, only a few low-mass protostars remain physically separated for the entire duration of the simulation. We note that this number depends sensitively on our choice of accretion radius, which might be too small considering that real protostars are surrounded by an unresolved accretion disk. Assuming a somewhat larger accretion radius would further reduce the number of low-mass protostars that could survive.}
\end{center}
\end{figure}

\section{Summary and Conclusions}

We have used the moving mesh code {\small AREPO} to follow the runaway collapse of the gas in five statistically independent minihalos from cosmological to protostellar scales -- over more than twenty orders of magnitude in density. We have captured the subsequent evolution of the newborn protostellar cloud for more than $100$ dynamical times with a sink-particle algorithm that resolves the accretion of the gas down to scales of $100\,{\rm R}_\odot$ or less, comparable to the maximum photospheric size of Pop~III protostars. As proposed by \citet{clark11b}, in all five minihalos a circumstellar disk forms that fragments vigorously into a small cluster of protostars with a range of masses. The gas becomes gravitationally unstable multiple times and continues to form protostars for the entire duration of the simulation. After only $1000\,{\rm yr}$ of accretion, of order $10$ protostars per minihalo have formed and display a relatively flat protostellar mass function ranging from $\sim 0.1$ to nearly $10\,{\rm M}_\odot$.

Although the sink-particle technique employed in this study allows a continuation of the calculations to much later times than was possible in previous {\it ab initio} simulations of primordial star formation \citep{abn02,yoh08}, a number of uncertainties remain. One problem is the artificially reduced density and pressure created by the removal of mesh-generating points around sink particles. However, we have found that using systematically smaller accretion radii results in more fragmentation and a decrease of the typical fragment mass, such that a correct treatment of the boundary region between sink particles and the gas should strengthen our conclusions. A more important caveat is the inability of the sink-particle algorithm to model the gasdynamical friction between physically extended protostars during close encounters. In an attempt to maximize this effect, we have implemented adhesive sink particles in addition to a more standard formulation of sink particles. This leads to an increased merger rate, so that fewer protostars survive. However, even this extreme assumption does not prevent the formation of a relatively rich cluster of protostars. A final caveat is that we do not self-consistently model the interaction of the protostars with the surrounding gas. It is unclear how important the resulting neglect of gasdynamical friction and torques is, since simulations that model the protostellar surface as well as the parent gas cloud are not yet feasible. However, it appears unlikely that this caveat will qualitatively affect our conclusions, since fragmentation typically occurs on scales larger than the minimum resolution length.

Modulo the uncertainties mentioned above, the simulations presented here portray a very different picture of primordial star formation than is commonly assumed. Instead of forming a single object, the gas in minihalos fragments vigorously into a number of protostars with a range of masses. It is an open question as to how this early mass function will be mapped into the final mass function of Pop~III, after accretion, fragmentation and merging have finally stopped. However, it is interesting to speculate how this nonlinear mapping will play out. If a flat, broad mass function persists, a number of lower-mass Pop III stars will have formed which could survive to the present day if their mass remains below $\sim 0.8\,M_\odot$. Although this possibility is speculative because of the above uncertainties and the fact that we follow the protostellar accretion only for the first $1000$ out of $10^5$ or $10^6\,{\rm yr}$, it is worth looking for such Pop~III fossils in ongoing and planned large surveys of metal-poor stars in the Milky Way \citep{bc05}. Specifically, such surveys should be focused on the Galactic bulge, where the Pop~III survivors should preferentially reside due to the biasing of the minihalo formation sites \citep{dmm05,gao10}. The planned Apache Point Observatory Galactic Evolution Experiment (APOGEE) with its near-IR capability may be well suited for this search \citep{majewski10}. Because of interstellar pollution, even true Pop~III fossils would appear as extreme Pop~II stars, but upper limits would be significantly lower than the currently probed values \citep{fjb09}.

Furthermore, the presence and mutual competition of multiple accretors in a given minihalo will act to limit the growth of the most massive objects. Pop~III stars are therefore less likely to reach masses in excess of $\sim 140\,{\rm M}_\odot$, the threshold for triggering extremely energetic pair-instability supernovae (PISNe) during stellar death \citep{hw02}. A reduced PISN rate is more easily compatible with the absence of their distinct nucleosynthetic signatures in any of the extremely metal-poor halo stars observed so far \citep{iwamoto05}. Pop~III stars could still have given rise to numerous extremely luminous supernova explosions, if they had masses of a few $10\,{\rm M}_\odot$ and if they were rapid rotators, as suggested by recent studies \citep{sbl11}. They would then have exploded as core-collapse hypernovae, with explosion energies that are similar to PISNe \citep{un02}. Finally, our results may challenge models of so-called `dark stars', which are Pop~III stars powered by DM self-annihilation heating \citep{freese08,iocco08}. These models invoke an increased DM interaction rate at the center of the Pop~III star which itself lies at rest at the center of its minihalo. On top of recent claims that the initial collapse is not substantially delayed by DM annihilations \citep{ripamonti10}, the complex dynamics of a protostellar cluster at the center of the minihalo may further preclude efficient DM capture and heating.

\acknowledgements{T.H.G. thanks Klaus Dolag for support on numerous technical issues, and Tom Abel, Jarrett Johnson and Rob Wiersma for many stimulating discussions. We thank the Leibniz Supercomputing Center of the Bavarian Academy of Sciences and Humanities, and the Computing Center of the Max-Planck-Society in Garching, where the simulations were carried out. R.J.S and R.S.K acknowledge subsidies from a Frontier grant of Heidelberg University sponsored by the German Excellence Initiative. R.S.K. acknowledges financial support from the {\em Baden-W\"{u}rttemberg Stiftung} via their program International Collaboration II (grant P-LS- SPII/18), from the German {\em Bundesministerium f{\"u}r Bildung und Forschung} via the ASTRONET project STAR FORMAT (grant 05A09VHA), and from the DFG under grants no. KL1358/1, KL1358/4, KL1358/5, KL1358/10, and KL1358/11. V.B. acknowledges support from NSF grants AST-0708795 and AST-1009928, as well as NASA through Astrophysics Theory and Fundamental Physics Program grants NNX 08-AL43G and 09-AJ33G.}


\end{document}